\documentclass[letterpaper, 10 pt, conference]{ieeeconf}  % Comment this line out if you need a4paper

\IEEEoverridecommandlockouts                              % This command is only needed if 
\usepackage{color}

\newtheorem{remark}{Remark}
\newtheorem{definition}{Definition}

\usepackage{amsmath}
\usepackage{graphicx}
\usepackage{bm}
\usepackage{amsfonts}
\usepackage{multirow}
\usepackage{arydshln}
\usepackage{cite}
\usepackage{float}

\let\labelindent\relax
\usepackage{enumitem}

\DeclareMathOperator*{\Min}{minimize}

\title{\LARGE \bf
Identifiability Analysis of a Pseudo-Two-Dimensional Model \& \\ 
Single Particle Model-Aided Parameter Estimation*,$\dagger$
}

\author{Luis D. Couto$^{1}$, Keivan Haghverdi$^{1,2}$, Feng Guo$^{1}$, Khiem Trad$^{1}$, Grietus Mulder$^{1}$ % <-this % stops a space
\thanks{*%This work was not supported by any organization
Part of this work has been done within the FLEXINet project which is being carried out with subsidy from the Dutch
Ministry of Economic Affairs and Climate Policy and the Ministry of the Interior and Kingdom Relations, under the
Mission-based Research, Development and Innovation (“MOOI regeling” in Dutch; project reference MOOI32027). 
And a part of this work has been done within the BE-CO-OPTIMAL project which is supported by the Energy Transition Fund of the Belgian Federal Public Service Economy.
}% <-this % stops a space
\thanks{$\dagger$%This is a preprint. The paper has been presented at the 2025 American Control Conference (ACC) and will appear in the conference proceedings. This version includes a supplementary table not present in the conference version due to page limits.
© 2025 IEEE. Personal use of this material is permitted. Permission from IEEE must be obtained for all other uses, including reprinting/republishing this material for advertising or promotional purposes, collecting new collected works for resale or redistribution to servers or lists, or reuse of any copyrighted component of this work in other works. }
\thanks{This is a preprint. The paper has been presented at the 2025 American Control Conference (ACC) and will appear in the conference proceedings. This version includes a supplementary table not present in the conference version due to page limits.
}
\thanks{$^{1}$Luis D. Couto, Keivan Haghverdi, Feng Guo, \mbox{Khiem Trad} and Grietus Mulder are with 
        Water \& Energy Transition unit,
        Vlaamse Instelling voor Technologisch Onderzoek (VITO), Boeretang 200, B-2400 Mol, Belgium
        {\tt\small \{luis.coutomendonca, keivan.haghverdi, feng.guo, khiem.trad, grietus.mulder\}@vito.be} \& EnergyVille, Thor park 8310, B-3600 Genk, Belgium.}%
\thanks{$^{2}$Keivan Haghverdi is also {with the Department of Chemistry, RWTH Aachen University, Germany
        {\tt\small keivan.haghverdi@rwth-aachen.de}}}%
}

\begin{document}

\maketitle
\thispagestyle{empty}
\pagestyle{empty}

% ri{This template proviaring electronic versions of their papers. All}

\begin{abstract}

This contribution presents a parameter identification methodology for the accurate and fast estimation of model parameters in a pseudo-two-dimensional (P2D) battery model. 
The methodology consists of three key elements. 
First, the data for identification is inspected and specific features herein that need to be captured are included in the model. 
Second, the P2D model is analyzed to assess the identifiability of the physical model parameters and propose alternative parameterizations that alleviate possible issues. 
Finally, diverse operating conditions are considered that excite distinct battery dynamics which allows the use of different low-order battery models accordingly. 
Results show that, under low current conditions, the use of low-order models achieve parameter estimates at least 500 times faster than using the P2D model at the expense of twice the error. 
However, if accuracy is a must, these estimated parameters can be used to initialize the P2D model and perform the identification in half of the time. 

% Accurate battery modeling is essential for optimizing the performance and lifespan of energy storage systems.
% Among the various physics-based battery models, the pseudo-two-dimensional (P2D) model is renowned for its accuracy, as it captures the intricate dynamics of electrochemical processes within a battery. However, its complexity results in significant computational demands, particularly during model identification where iterative optimization of numerous parameters can be prohibitively time-consuming. 
% To address this challenge, we propose a novel approach that leverages the efficiency of a single particle model (SPM) for parameter identification. The SPM, with its reduced complexity and faster computation time, is utilized to estimate the battery parameters efficiently. These parameters are then employed in the more accurate but computationally intensive P2D model. By integrating the rapid SPM for initial parameter estimation, our method significantly reduces the computational burden associated with P2D model parameter optimization, thereby enhancing overall efficiency and enabling more practical applications of high-fidelity battery models.

\end{abstract}

%%%%%%%%%%%%%%%%%%%%%%%%%%%%%%%%%%%%%%%%%%%%%%%%%%%%%%%%%%%%%%%%%%%%%%%%%%%%%%%%
%\vspace{-0.1cm}

\section{INTRODUCTION}

%\vspace{-0.1cm}

%\gi{general intro, P2D and SPM:} 
The increasing reliance on renewable energy sources and the surge in electric vehicle adoption have significantly heightened the importance of lithium-ion batteries. To design and operate these batteries efficiently, employing physics-based battery modeling has become essential. One such model is the pseudo-two-dimensional (P2D) battery model, introduced by Doyle, Fuller, and Newman \cite{doyle1993modeling,fuller1994simulation}. %,newman1962theoretical,newman1975porous}. 
While this model provides a comprehensive representation of battery behavior, its complexity can adversely affect computational efficiency and speed. 
%\gi{reduced-order models:} 
Numerous efforts have been made to simplify the P2D model, leading to the development of various reduced-order models. Among these, the single particle model (SPM) is one of the most widely recognized and utilized \cite{%atlung1979dynamic,
%di2010lithium,
santhanagopalan2006online,moura2014adaptive}. %,ning2004cycle}. 
However, the SPM is primarily effective in low current regions, and its accuracy tends to diminish at high C-rates \cite{chaturvedi2010modeling}.

%\gi{model parameters and identifiability:} 
Physics-based models are inherently complex and often involve a substantial number of parameters. For example, the P2D model typically includes %between 15 and 
30 physical parameters whereas the SPM generally has %9 
14 parameters \cite{drummond2020structural,bizeray2018identifiability}. 
%Sensitivity analysis can help to reduce this number by identifying which parameters have minimal impact on the model's output, allowing us to filter out less significant ones. However, even with this reduction, the remaining parameters can still pose significant challenges. 
%\gi{identifiability:} 
% In particular, the parameters in the P2D model and the SPM present difficulties for identification through non-destructive methods alone \cite{p2dunindent, bizeray2018identifiability}. This is due to the mathematical interdependence of these parameters, which often appear as products in equations, making it challenging to isolate their individual effects \cite{couto2023lithium}. 
In particular, many parameters in these models require destructive measurements by physically opening the battery to obtain them, which is expensive, labour intensive and time consuming. %and sometimes insufficiently precise. 
Moreover, many of these parameters cannot be directly measured, necessitating the use of parameter estimation methods \cite{p2dunindent, bizeray2018identifiability}. 
However, even parameter estimation methods face challenges in accurately estimating many parameters, as they are mathematically interdependent and often appear as products in equations, making it difficult to isolate their individual effects \cite{couto2023lithium}.
One way to address this is by grouping parameters, which renders P2D models of 21 to 24 grouped parameters \cite{drummond2020structural,Khalik-2021} and SPMs of 7 grouped parameters \cite{bizeray2018identifiability,Marquis-2019}. 
It should be kept in mind that each model with grouped parameters reported in the literature has its own underlying assumptions, making it difficult a direct comparison. 
Once the grouped parameters have been identified from data, some of the intertwined physical parameters can then be derived if some other physical quantities have been determined independently. %independently through alternative methods, such as dedicated laboratory experiments. %Unfortunately, these experiments can be expensive, labour intensive, time consuming and sometimes insufficiently precise. 

% \ri{more on grouping?}

%\gi{parameter ID:} 
For the identification of parameters 
%To address the challenges of identifying parameters 
in physics-based models, it is common practice to use parameter optimization techniques \cite{fan2020systematic,rahman2016electrochemical,pang2019parameter}. %,jokar2016inverse}. 
These methods seek to find the set of parameters that best align the model's predictions with experimental data. However, this process can be particularly time-consuming for models like the P2D, known for its slow computational speed, as the model must be executed multiple times during the optimization procedure. Consequently, the literature reports extensive computation times, often ranging from hours to weeks, depending on the model's complexity and the optimization method used \cite{forman2012genetic, zeng2019global, deng2021sensitivity}. 
Hybrid minimization algorithms have also been developed to accelerate the parameter estimation process of a P2D model in \cite{Reddy-2019} considering 44 physical parameters, taking 14 h and reaching 7 mV of mean absolute error.

%\ri{accelerate param ID?}
%\ri{compare your norm with others like Ross and Thijs?}

% Given these constraints, this research aims to utilize a reduced-order model, such as the SPM, for the initial parameter identification. By leveraging the faster computational speed of the SPM, we can efficiently determine a set of parameters, which can then be used to optimize the P2D model. This approach is intended to streamline the parameter identification process and reduce overall computational time.

%\gi{in this work...:} 
In this research (to appear in \cite{Couto-2025}), we have detected several gaps in the literature that we attempted to fill. 
First, we account for specific features evidenced in the data like CV operation and thermal dynamics. 
The former is rarely discussed in the literature where the focus is put on current-controlled data (e.g. \cite{fan2020systematic,pang2019parameter}). 
%Thermal dynamics %for high demand operation, 
The latter are necessary for accurate modeling during operating conditions with high demand of current. %over a wide range of operating conditions. 
Thus, a wide operating range (from ${\rm C}/20$ to $3{\rm C}$) is considered here. 
Second, an identifiability analysis is performed over the P2D model in order to show the pitfalls of using an as-made model and the necessity of its normalization to avoid overparameterization and alleviate non-identifiability. 
Finally, we derive an identification methodology for an efficient and accurate parameter identification of a P2D model that leverages an SPM when low currents are used and avoids the hard process of manual tuning. 
The identified SPM parameters are obtained quickly and they can be used as a final result, or to initialize the parameter identification of a P2D model if higher accuracy is required. 
The proposed cross-model methodology is intended to streamline the parameter identification process of high-order models while reducing the overall computational time for a given acceptable error level. 
This is in contrast with classical methodologies where a single model is considered (c.f. the review paper \cite{miguel2021review} that examined over 30 studies on parameter identification using physics-based models).

% \ri{Other key insights that can be highlighted include:}
% \begin{itemize}
%     % \item \gi{CV modeling is rarely discussed in the literature, with most studies focusing primarily on the CC protocol.}
%     % \item \gi{Due to the high number of parameters, manual tuning of the parameters is an extremely hard and time-consuming process}
%     \item \gi{ Most studies focus on parameter identification within a single model. For example, in a review paper by Miguel et al. \cite{miguel2021review}, which examined over 30 studies on parameter identification using physics based models, we found that none of the cases employed a cross-model parameter identification strategy.
%    However, some studies have attempted to connect the parameters of the equivalent circuit model with physics-based models \cite{geng2021bridging,li2019physics}.}
% \end{itemize}

%\clearpage

%\vspace{-0.1cm}

\section{PROBLEM FORMULATION}

%\vspace{-0.1cm}

\begin{figure}[!htb]
	\centering
	\includegraphics[scale=0.55,trim={4cm 13cm 8.7cm 1.5cm},clip]{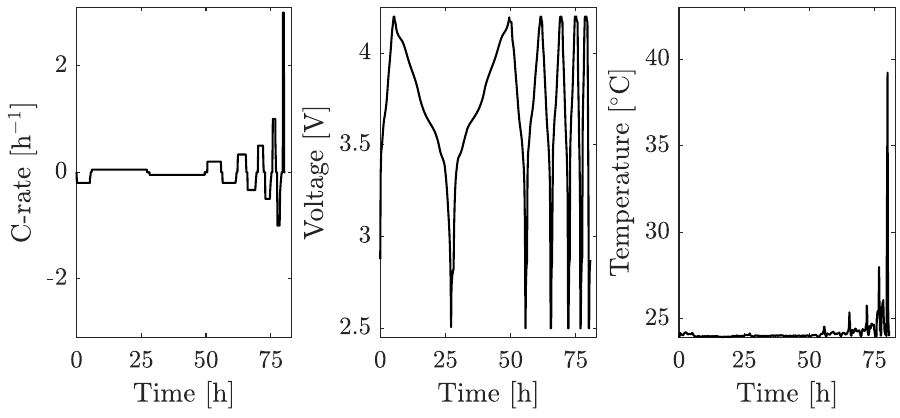}
	\vspace{-0.4cm}
	\caption{Current, voltage and temperature measurements from a lithium-ion battery cell subjected to CCCV charge-discharge cycles from ${\rm C}/20$ to $3{\rm C}$.}
	\label{f1}
\vspace{-0.2cm}
\end{figure}

Our final goal is to parameterize a P2D model with the data set $\mathcal{D}$ shown in Fig. \ref{f1}. 
%This dataset is denoted as $\mathcal{D} = \{(i(t_k), v_{\rm exp}(t_k), T(t_k)), t_k = 1,\ldots,N_t\}$, where $v_{\rm exp}(t_k)$ and $T(t_k)$ are respectively the voltage and surface temperature sequences associated to the current sequence $i(t_k)$ for a given sampled time $t_k$ and data samples $N_t$. 
This data consists of a series of charge-discharge cycles that were obtained from constant current-constant voltage (CCCV) protocol with 30 min of rest between each non-zero current step with the considered current magnitudes $\{{\rm C}/20, {\rm C}/5, {\rm C}/3, {\rm C}/2, 1{\rm C}, 3{\rm C}\}$.
\vspace{-0.2cm}

In order to identify the P2D model parameters, the following outstanding assumptions are considered:
\begin{enumerate}[label={}, leftmargin=1cm]
\item[($\mathcal{A}$1)] The cell chemistry is known, and nominal model parameters are either taken from the literature or experimentally determined.
\item[($\mathcal{A}$2)] The positive and negative electrode OCPs are measured via half cells and a function is fitted to them.
\end{enumerate}

Then, the model parameters identification problem can be cast as a generic optimization problem of the form
\vspace{-0.2cm}
% \begin{eqnarray} 
\begin{subequations}
\label{opt_gen0}
\begin{align}
&\Min_{{\theta}} \hspace{0.2cm} %\sqrt{\frac{1}{N} 
\sum_{t=0}^{N} w(t) \left( v_{\rm exp}(t) - v(t,{\theta}) \right)^2%} 
\label{eq:opt_gen1} \\
&\hspace{0.65cm}\mathrm{subject\ to\ model\ dynamics,} %\ \eqref{eq:diff_paper},\eqref{eq:out_norm},
\label{eq:st_opt_gen1} %\\
%&\hspace{2.33cm}\mathrm{parameter\ bounds,} %\ \eqref{eq:diff_paper},\eqref{eq:out_norm},
%\label{eq:st_opt_gen2}
%\vspace{-0.2cm}
\end{align}
\end{subequations}
% \end{eqnarray}
where $t$ stands for time, %is the discrete-time variable, 
$v_{\rm exp}$ is the measured voltage, $v$ is the modeled voltage, ${\theta}$ is the parameter vector and $w$ is a weight function.

% Several challenges can be distinguished when estimating P2D model parameters for the considered data. Among them, the most relevant ones are:
% \begin{enumerate}
% \item The considered dataset covers a wide range of operating conditions from low (${\rm C}/20$) to high (${\rm 3}{\rm C}$) currents.
% \item Due to the previous aspect, cell temperature changes drastically.
% \item The data involves CV charge and discharge steps besides classical CC.
% \item The P2D model is over-parameterized and possibly non-identifiable.
% \item The complexity of such model makes it to incur in a high computational burden.
% \end{enumerate}
% These challenges are tackled by the proposed identification methodology presented in the following sections.

The problem \eqref{opt_gen0} is solved via electrochemical modeling and parameter identification as presented in the following sections.

%\clearpage
%\vspace{-0.1cm}

\section{BATTERY MODELING}

%\vspace{-0.1cm}

In this section, the P2D model as well as the SPM are described. Moreover, special features that are in the data and require modeling are also introduced, followed by a normalized version of the considered models.

%\vspace{-0.2cm}

\subsection{Pseudo-Two-Dimensional Model}

%\vspace{-0.1cm}

The P2D model is one of the most widely used electrochemical battery models for simulating battery behavior \cite{doyle1993modeling, fuller1994simulation}. %, newman1962theoretical, newman1975porous}. 
This model consists of a set of partial differential equations (PDEs) with algebraic constraints, which are defined along the $x$-axis spanning the length of the battery. Additionally, the active particles within the electrodes are modeled as spheres, with their behavior described along the radial coordinate $r$. The original P2D model equations with their respective boundary conditions and additional equations can be found in \cite{doyle1993modeling, fuller1994simulation}. 
They are also reported in the appendix in Table \ref{t:p2d_app}, along with the description of all the variables and parameters in Table \ref{t:phys_params}, for the sake of completeness. 
%and not shown here due to space constraints
%\footnote{The normalized version of the P2D model proposed here is shown below in Table \ref{t:p2d_norm}. It is considered in this section just to have a reference of model equations, but the as-made P2D model \cite{doyle1993modeling,fuller1994simulation} is the one referred to here.}.
%are shown in Table \ref{t:p2d_OG}, whereas the description of all the variables and parameters are reported in Table \ref{t:phys_params} in the appendix section. 
The main equations are:

\begin{itemize}[leftmargin=*]
    \item Solid-phase diffusion: %: equation \eqref{spd} 
    equation \eqref{spd_new_app} describes the diffusion of lithium within the electrode's solid particles, governed by Fick's law. It models the change in lithium concentration over time and space within the electrode particles.
    \item Kinetics: equation \eqref{kn_new_app}, %: equation \eqref{kn}, 
    often represented by the Butler-Volmer equation, models the electrochemical reactions at the electrode-electrolyte interface. It relates current density to overpotential, covering both anodic and cathodic reactions.    
    \item Mass Balance: equation \eqref{mb_new_app} ensures the conservation of lithium ions in the electrolyte, accounting for both diffusion and migration due to concentration gradients and electric fields. It accurately reflects ion distribution, impacting internal resistance and performance.    
    \item Potential in solution: equation \eqref{psolution_new_app} describes the potential that is influenced by both the conductivity of the electrolyte and the overpotential arising from concentration gradients. 
    This potential is often described by incorporating the Nernst potential, which adjusts the particle open-circuit potential (OCP) to account for variations in lithium-ion concentration within the electrolyte.    
    \item Potential in solid: equation \eqref{psolid_new_app} governs the potential distribution within the solid electrode phase using Ohm's law. It relates current density to the solid-phase potential gradient.    
    \item Voltage: equation \eqref{volt_new_app} determines the terminal voltage of the battery as the difference between the positive and negative electrode solid potentials minus other potential drops caused by e.g. a film resistance in an electrode. %and resistive losses in the electrolyte. This equation provides the measurable voltage across the battery terminals.
\end{itemize}

%%------------------------------------------------
\begin{table*}[!htb]
 \vspace{0.2cm}
	\caption{Normalized P2D model equations with grouped and physical parameters defined in Table \ref{t:grouped_params} and Table \ref{t:phys_params}, respectively.}
 \vspace{-0.3cm}
	\centering
	
{\footnotesize
%{\small
\begin{tabular}{l c c c}
\hline
\textbf{Name} \hspace{-0.7cm} & \textbf{Equation} &%\hspace{2cm} 
\textbf{Boundary conditions/additional equations} &
\hspace{-1cm} \textbf{Eq.}\\ 
\hline
\begin{minipage}{1.5cm}
Solid-phase\\
diffusion
\end{minipage}
 \hspace{-0.5cm}
&
\parbox{7cm}{
\begin{equation} 
% \displaystyle \frac{\partial {c}_{s}^\pm}{\partial t} = 
% \frac{1}{r^2} \frac{\partial}{\partial r} 
% \left( r^2 D_s \frac{\partial c_s}{\partial r} \right)
\displaystyle \frac{\partial \bar{c}_{s}^m}{\partial t} (\bar{x}^m\!\!,\bar{r},t) = 
\frac{1}{\tau_{d,s}^m} 
% \left( \frac{2}{\bar{r}} \frac{\partial \bar{c}_{s}^\pm}{\partial \bar{r}}(\bar{x},\bar{r},t) + 
% \frac{\partial^2 \bar{c}_{s}^\pm}{\partial \bar{r}^2}(\bar{x},\bar{r},t) \right)
\frac{1}{\bar{r}^2} \frac{\partial}{\partial \bar{r}} \left( \! \bar{r}^2 \frac{\partial \bar{c}_s^m}{\partial \bar{r}} (\bar{x}^m\!\!,\bar{r},t) \! \right)
\nonumber
\end{equation}
}
& \hspace{-0.2cm}
\parbox{8.2cm}{
% \begin{eqnarray} 
\begin{subequations}
\label{spd_new}
\begin{gather}
% \displaystyle \left. \frac{\partial \bar{c}_s^\pm}{\partial \bar{r}} \right\vert_{\bar{r}=0} &\!\!\!\!\!\!\!\! = &\!\!\!\! 0, \;
\displaystyle 
\left. \frac{\partial \bar{c}_s^m}{\partial \bar{r}}(\bar{x}^m\!\!,\bar{r},t) \right\vert_{\bar{r}=0} \!\!\!\!\!\!\!\!= 0, \;
\displaystyle  
 \left. \frac{\partial \bar{c}_s^m}{\partial \bar{r}}(\bar{x}^m\!\!,\bar{r},t) \right\vert_{\bar{r}=1} \!\!\!\!\!\!\!\!= -\frac{\tau_{d,s}^m}{3 %q_s^m
 \tau_{c,s}^m} \tilde{j}_n^m(\bar{x}^m\!\!,t) 
 % \nonumber
\label{spd1_new}
  \\
 %\bar{c}_{s,{\rm avg}}^m ({x},t) = \displaystyle -\frac{1}{q_s^m} i(t) 
 \bar{c}_{s,{\rm avg}}^m (\bar{x}^m\!\!,t) = \displaystyle 3 \int_0^{1} \bar{r}^2 \bar{c}_{s} (\bar{x}^m\!\!,r,t) {\rm d}\bar{r} 
 % \nonumber
\label{spd2_new}
 \\
 \bar{c}_{ss}^\pm (\bar{x}^m\!\!,t) = \left. \bar{c}_{s}^\pm (\bar{x}^m\!\!,\bar{r},t) \right\vert_{\bar{r}=1}
 % \nonumber
\label{spd3_new}
% \label{spd}
%\nonumber
% \\
% \displaystyle  
%  \left. \frac{\partial \bar{c}_s^\pm}{\partial \bar{r}} \right\vert_{\bar{r}=1} &\!\!\!\!\!\!\!\!= &\!\!\!\! -\frac{\tau_{s}^\pm}{3 q^\pm} J_n^\pm
% \nonumber
\end{gather}
\end{subequations}
% \end{eqnarray}
}
% & %\hspace{1.5cm}
% \parbox{0.5cm}{
% \begin{eqnarray} 
% \hspace{-0.3cm}
% \label{spd_new}
% \end{eqnarray}
% }
\vspace{-0.3cm}
\\
\begin{minipage}{1.5cm}
Kinetics$^a$
\end{minipage}
 \hspace{-0.7cm}
&
\parbox{7cm}{
\begin{eqnarray} 
%    \tilde{j}_n^m(\bar{x}^m,t) &\!\!\!\!=&\!\!\!\! \tilde{i}_0^m (\bar{x}^m,t) \left( \exp \left( \beta \eta^m (\bar{x}^m,t) \right) - \exp \left( -\beta \eta^m (\bar{x}^m,t) \right) \right)
    \tilde{j}_n^m(\bar{x}^m\!\!,t) &\!\!\!\!=&\!\!\!\! \tilde{i}_0^m (\bar{x}^m\!\!,t) \left( e^{ \displaystyle \beta \eta^m (\bar{x}^m\!\!,t) } - e^{ \displaystyle -\beta \eta^m (\bar{x}^m\!\!,t) } \right)
    \nonumber
    % \\
    % I_0^\pm &\!\!\!\!=&\!\!\!\! 3 \tau_{k}^\pm q_{s}^\pm \sqrt{\bar{c}_e \bar{c}_{s}^\pm (1-\bar{c}_{s}^\pm)}
    % \nonumber
    % \\
    % \eta_s^\pm &\!\!\!\!=&\!\!\!\! \phi_{s}^\pm - \phi_{e} - U_s^\pm 
    % \nonumber
\end{eqnarray}
}
& \hspace{-0.2cm}
\parbox{8.2cm}{
% \begin{eqnarray} 
\begin{subequations}
\label{kn_new}
\begin{gather}
    \tilde{i}_0^m (\bar{x}^m\!\!,t) = %3 %q_{s}^m
%    \tau_{c,s}^m 
	\frac{3 \tau_{c,s}^m }{\tau_{k}^m}
	\sqrt{\bar{c}_e^m (\bar{x}^m\!\!,t) \bar{c}_{ss}^m (\bar{x}^m\!\!,t) (1 - \bar{c}_{ss}^m (\bar{x}^m\!\!,t))} %/ \tau_{k}^m
    % \nonumber
    \label{kn1_new}
    \\
    \eta^m(\bar{x}^m\!\!,t) = \phi_{s}^m (\bar{x}^m\!\!,t) - \phi_{e}^m (\bar{x}^m\!\!,t) - U^m (\bar{x}^m\!\!,t) 
    % \nonumber
    \label{kn2_new}
% \end{eqnarray}
\end{gather}
\end{subequations}
}
% & %\hspace{1.5cm}
% \parbox{0.5cm}{
% \begin{eqnarray} 
% \hspace{-0.3cm}
% \label{kn_new}
% \end{eqnarray}
% }
\vspace{-0.3cm}
\\
\begin{minipage}{1.5cm}
Mass\\
balance %$^b$
\end{minipage}
 \hspace{-0.7cm}
&
\parbox{7cm}{
\begin{equation} 
    \displaystyle \frac{\partial \bar{c}_{e}^{m}}{\partial t} (\bar{x}^m\!\!,t) = \frac{1}{\tau_{d,e}^{m}}  \frac{\partial^2 \bar{c}_{e}^{m}}{\partial (\bar{x}^m)^2} (\bar{x}^m\!\!,t) + \frac{\gamma}{%q_e^{m}
    {\nu_{e}^{m}}} \tilde{j}_n^{m} (\bar{x}^m\!\!,t)
    \nonumber
\end{equation}
}
& \hspace{-0.2cm}
\parbox{8.2cm}{
% \begin{eqnarray} 
\begin{subequations}
\label{mb_new}
\begin{gather}
\displaystyle %\frac{\varepsilon_e^-}
% \frac{1}{\tau_{e}^-}
\frac{{\nu_e^-}}{{\tau_{d,e}^-}} \left. \frac{\partial \bar{c}_{e}^-}{\partial \bar{x}^-} (\bar{x}^-\!\!,t) \right\vert_{\bar{x}^-=0} \!\!\!\!\!\!\!\!=
\displaystyle %\frac{\varepsilon_e^+}
% \frac{1}{\tau_{e}^+} 
\frac{{\nu_e^+}}{{\tau_{d,e}^+}} \left. \frac{\partial \bar{c}_{e}^+}{\partial \bar{x}^+} (\bar{x}^+\!\!,t) \right\vert_{\bar{x}^+=1} \!\!\!\!\!\!\!\!\!\!= 0, \;\;
% \displaystyle %\frac{\varepsilon_e^-}{\tau_{e}^-}
% % \frac{\varepsilon_e^{-s}}{\tau_{e}^-} 
% \frac{\mi{\nu_e^-}}{\mi{\tau_{d,e}^-}} \left. \frac{\partial \bar{c}_{e}^-}{\partial \bar{x}} (\bar{x},t) \right\vert_{\bar{x} = \bar{\delta}^-} \!\!\!\!\!\!\!\!=...
% \nonumber
\label{mb1_new}
\\
\displaystyle 
\frac{{\nu_e^-}}{{\tau_{d,e}^-}} \left. \frac{\partial \bar{c}_{e}^-}{\partial \bar{x}^-} (\bar{x}^-\!\!,t) \right\vert_{\bar{x}^- = 1} 
\!\!\!\!\!\!\!\!=
\frac{{\nu_e^s}}{{\tau_{d,e}^s}} \left. \frac{\partial \bar{c}_{e}^s}{\partial \bar{x}^s} (\bar{x}^s\!\!,t) \right\vert_{\bar{x}^s = 0}, \;\;
% \nonumber
\label{mb12_new}
\\
% \displaystyle \frac{1}{\tau_{e}^-} \left. \frac{\partial \bar{c}_{e}^-}{\partial \bar{x}^-} \right\vert_{\bar{x}^- = 1} &\!\!\!\!\!\!\!\!\!\!\!\!\!\!= &\!\!\!\!\!\! 
%  \frac{1}{\tau_{e}^s} \left. \frac{\partial \bar{c}_{e}^s}{\partial \bar{x}^s} \right\vert_{\bar{x}^s = 0}
% \nonumber 
% \\
% \displaystyle ...= %\frac{\varepsilon_e^s}
% % \frac{1}{\tau_{e}^s}
% \frac{\mi{\nu_e^s}}{\mi{\tau_{d,e}^s}} \left. \frac{\partial \bar{c}_{e}^s}{\partial \bar{x}} (\bar{x},t) \right\vert_{\bar{x} = \bar{\delta}^-}, \;\;
%\frac{\varepsilon_e^s}
% \frac{\varepsilon_e^{s+}}{\tau_{e}^s}
\frac{{\nu_e^s}}{{\tau_{d,e}^s}} \left. \frac{\partial \bar{c}_{e}^s}{\partial \bar{x}^s} (\bar{x}^s\!\!,t) \right\vert_{\bar{x}^s = 1} \!\!\!\!\!\!\!\! =
 %\frac{\varepsilon_e^+}
 % \frac{1}{\tau_{e}^+}
\frac{{\nu_e^+}}{{\tau_{d,e}^+}} \left. \frac{\partial \bar{c}_{e}^+}{\partial \bar{x}^+} (\bar{x}^+\!\!,t) \right\vert_{\bar{x}^+ = 0}
\label{mb2_new}
% \nonumber
% \\
% \displaystyle \frac{1}{\tau_{e}^+} \left. \frac{\partial \bar{c}_{e}^+}{\partial \bar{x}^+} \right\vert_{\bar{x}^+=1} &\!\!\!\!\!\!\!\!\!\!\!\!\!\!= &\!\!\!\!\!\! 
% 0
% \nonumber
% \end{eqnarray}
\end{gather}
\end{subequations}
}
% & %\hspace{1.5cm}
% \parbox{0.5cm}{
% \begin{eqnarray} 
% \hspace{-0.3cm}
% \label{mb_new}
% \end{eqnarray}
% }
\vspace{-0.3cm}
\\
\begin{minipage}{1.5cm}
Potential in \\
solution
\end{minipage}
 \hspace{-0.7cm}
&
\parbox{7cm}{
\begin{equation} 
    \tilde{i}_e^m (\bar{x}^m\!\!,t) = -{\kappa}^m \frac{\partial \phi_{e}^{m} }{\partial \bar{x}^m} (\bar{x}^m\!\!,t) + \frac{{\kappa}^{m} }{\beta} %\gamma 
    k_{\gamma f} \frac{\partial \ln c_{e}^{m} }{\partial \bar{x}^m} (\bar{x}^m\!\!,t)
    \nonumber
\end{equation}
}
& \hspace{-0.2cm}
\parbox{8.2cm}{
% \begin{eqnarray} 
\begin{subequations}
\label{psolution_new}
\begin{gather}
\displaystyle \left. \tilde{i}_{e}^-(\bar{x}^-\!\!,t) \right\vert_{\bar{x}^-=0} 
= 
\displaystyle \left. \tilde{i}_{e}^+(\bar{x}^+\!\!,t) \right\vert_{\bar{x}^+=1}
= 
0, 
% \nonumber
\label{psolution1_new}
\vspace{-0.3cm}
\\
\displaystyle \left. \tilde{i}_{e}^-(\bar{x}^-\!\!,t) \right\vert_{\bar{x}^-=1} 
= 
\displaystyle \left. \tilde{i}_{e}^s(\bar{x}^s\!\!,t) \right\vert_{\bar{x}^s = 1}
 =
\tilde{i}(t), 
% \nonumber
\label{psolution2_new}
% \\
% \displaystyle \left. I_{e}^s \right\vert_{\bar{x}^s = 1} &\!\!\!\!\!\!\!\!= &\!\!\!\! \bar{I}, 
% \nonumber
% \\
% \displaystyle \left. I_{e}^+ \right\vert_{\bar{x}^+=1} &\!\!\!\!\!\!\!\!= &\!\!\!\! 0, 
% \nonumber
% \end{eqnarray}
\end{gather}
\end{subequations}
}
% & %\hspace{1.5cm}
% \parbox{0.5cm}{
% \begin{eqnarray} 
% \hspace{-0.3cm}
% \label{psolution_new}
% \end{eqnarray}
% }
\vspace{-0.3cm}
\\
\begin{minipage}{1.5cm}
Potential in\\
solid
\end{minipage}
 \hspace{-0.7cm}
&
\parbox{7cm}{
\begin{eqnarray} 
    \tilde{i}_{s}^{m} (\bar{x}^m\!\!,t) &\!\!\!\!\!\!=&\!\!\!\!\!\! -\sigma^{m} \frac{\partial \phi_{s}^{m}}{\partial \bar{x}^m} (\bar{x}^m\!\!,t), \;
    \tilde{j}_n^{m}(\bar{x}^m\!\!,t) = \frac{\partial \tilde{i}_{e}^{m}}{\partial \bar{x}^m} (\bar{x}^m\!\!,t), 
    \nonumber 
    \\
    \tilde{i}(t) &\!\!\!\!\!\!=&\!\!\!\!\!\! \tilde{i}_{s}^{m}(\bar{x}^m\!\!,t) + \tilde{i}_{e}^{m}(\bar{x}^m\!\!,t)
    \nonumber 
    % \\
    % \bar{I} &\!\!\!\!=&\!\!\!\! I_{s} + I_{e}
    % \nonumber
    % \\
    % J_n &\!\!\!\!=&\!\!\!\! \frac{\partial I_{e}}{\partial \bar{x}}
    % \nonumber
\end{eqnarray} 
}
& \hspace{-0.2cm}
\parbox{8.2cm}{
% \begin{eqnarray} 
\begin{subequations}
\label{psolid_new}
\begin{gather}
\displaystyle \left. \tilde{i}_{s}^- (\bar{x}^-\!\!,t) \right\vert_{\bar{x}^-=0} 
=
\displaystyle \left. \tilde{i}_{s}^+ (\bar{x}^+\!\!,t) \right\vert_{\bar{x}^+=1}
=
\tilde{i}(t), 
% \nonumber
\label{psolid1_new}
\\
\displaystyle \left. \tilde{i}_{s}^- (\bar{x}^-\!\!,t) \right\vert_{\bar{x}^-=1} 
=
\displaystyle \left. \tilde{i}_{s}^+ (\bar{x}^s\!\!,t) \right\vert_{\bar{x}^s=1} 
=
0, 
\label{psolid2_new}
% \nonumber
% \\
% \displaystyle \left. I_{s} \right\vert_{\bar{x}^s=1} &\!\!\!\!\!\!\!\!= &\!\!\!\! 0, 
% \nonumber
% \\
% \displaystyle \left. I_{s} \right\vert_{\bar{x}^+=1} &\!\!\!\!\!\!\!\!= &\!\!\!\! \bar{I}, 
% \nonumber
% \end{eqnarray}
\end{gather}
\end{subequations}
}
% & %\hspace{1.5cm}
% \parbox{0.5cm}{
% \begin{eqnarray} 
% \hspace{-0.3cm}
% \label{psolid_new}
% \end{eqnarray}
% }
\vspace{-0.3cm}
\\
% \begin{minipage}{1.5cm}
% Lithium moles
% \end{minipage}
% &
% \parbox{7cm}{
% \begin{equation} 
%     \ri{\tau_{n_{Li}}(t)} = \tau_{c,s}^+ \bar{c}_{s,{\rm avg}}^+(t) + \tau_{c,s}^- \bar{c}_{s,{\rm avg}}^-(t) 
%     \nonumber
% \end{equation}
% }
% & %\hspace{1.5cm}
% \parbox{8cm}{
% % \begin{equation} 
% \begin{subequations}
% \label{nli_new}
% \begin{gather}
% -
% % \nonumber
% \label{nli1_new}
% % \end{equation}
% \end{gather}
% \end{subequations}
% }
% %& %\hspace{1.5cm}
% % \parbox{0.5cm}{
% % \begin{eqnarray} 
% % \hspace{-0.3cm}
% % \label{volt}
% % \end{eqnarray}
% % }
% \vspace{-0.3cm}
% \\
\begin{minipage}{1.5cm}
Voltage
\end{minipage}
 \hspace{-0.7cm}
&
\parbox{7cm}{
\begin{equation} 
    v(t) = \left. \phi_{s}^+ (\bar{x}^+,t) \right\vert_{x^+=1} - \left. \phi_{s}^- (\bar{x}^-,t) \right\vert_{x^-=0}  - r_{f} \tilde{i}(t)
    \nonumber
\end{equation}
}
& \hspace{-0.2cm}
\parbox{8.2cm}{
% \begin{equation} 
\begin{subequations}
\label{volt_new}
\begin{gather}
-
\label{volt1_new}
% \nonumber
% \end{equation}
\end{gather}
\end{subequations}
}
% & %\hspace{1.5cm}
% \parbox{0.5cm}{
% \begin{eqnarray} 
% \hspace{-0.3cm}
% \label{volt_new}
% \end{eqnarray}
% }
\vspace{-0.1cm}
\\
\hline
\end{tabular}
\\
$^a$Assuming $\alpha = 1 - \alpha = 1/2$.\\
% $^b$Normalized variables $\bar{\delta}^- = \delta^- / \ell$ and $\bar{\delta}^s = \delta^s / \ell$.\\
} 
% \\
% \begin{tabular}{l}
% %{\scriptsize $^a$Assuming $\alpha = 1/2$.} \\
% \end{tabular}

	\label{t:p2d_norm}
%\vspace{-0.6cm}
\end{table*}
%%------------------------------------------------

%%------------------------------------------------
\begin{table}[!htb]
\vspace{0.3cm}
	\caption{Grouped parameters of P2D model in Table \ref{t:p2d_norm}.}
    \vspace{-0.3cm}
	\centering

\begin{tabular}{l c c}
\hline
Definition \hspace{-0.6cm} & Symbol \hspace{-0.6cm}& Expression \\ 
\hline
\multicolumn{3}{c}{Main grouped parameters} \\
{Solid diffusion time constant} [s] \hspace{-0.6cm}
& %\hspace{-1cm}
$\tau_{d,s}$ \hspace{-0.6cm}
&
$R_s^2 / D_s$
\vspace{0.1cm}
\\
% {Electrode capacity} [C] \hspace{-0.5cm}
{Electrode charge time} [s] \hspace{-0.6cm}
& %\hspace{-1cm}
% $q_s$ \hspace{-0.5cm}
$\tau_{c,s}$ \hspace{-0.6cm}
&
$F \varepsilon_s L A c_{s,{\rm max}} / i_{\rm ref}$
\vspace{0.1cm}
\\
{Reaction time scale} [s] \hspace{-0.6cm}
& %\hspace{-1cm}
$\tau_{k}$ \hspace{-0.6cm}
&
% $k_n \sqrt{c_{e}^0} / R_s$
$R_s / (k_n \sqrt{c_{e,{\rm ref}}})$
\vspace{0.1cm}
\\
Elyte. diffusion time constant [s] \hspace{-0.6cm}
& %\hspace{-1cm}
$\tau_{d,e}$ \hspace{-0.6cm}
&
$%\ell^2 
L^2 / (\varepsilon_e^{-1} D_{e,{\rm eff}})$
\vspace{0.1cm}
\\
% {Electrolyte "speed" constant}$^a$ [m$\cdot$s$^{-1}$] \hspace{-0.5cm}
% & %\hspace{-1cm}
% $\nu_{e}^{-1}$ \hspace{-0.5cm}
% &
% $D_{e,{\rm eff}} / \ell$
% \vspace{0.1cm}
% \\
% {Electrolyte "capacity"}$^a$ [C] \hspace{-0.5cm}
% {Electrolyte "charge time"}$^a$ [s] \hspace{-0.5cm}
% Elyte. "input coeff."$^a$ [mol$\cdot$m$^{-2}\cdot$A$^{-1}$] \hspace{-0.5cm}
{Elyte. "input coeff."$^a$ [s$\cdot$m$^{-2}$]} \hspace{-0.6cm}
& %\hspace{-1cm}
% $q_e$ \hspace{-0.5cm}
$\nu_{e}$ \hspace{-0.6cm}
&
% $F \varepsilon_e L A c_{e,{\rm ref}}  / i_{\rm ref}$
% $\varepsilon_e L c_{e,{\rm ref}}  / i_{\rm ref}$
{$F \varepsilon_e L c_{e,{\rm ref}}  / i_{\rm ref}$}
\vspace{0.1cm}
\\
% {Ionic conductivity} [S] \hspace{-0.5cm}
{Ionic cond. per unit current} [V$^{-1}$] \hspace{-0.6cm}
& %\hspace{-1cm}
$\kappa$ \hspace{-0.6cm}
&
$A \kappa_{e,{\rm eff}} / (L i_{\rm ref})$
\vspace{0.1cm}
\\
% {Electronic conductivity} [S] \hspace{-0.5cm}
{Elect. cond. per unit current} [V$^{-1}$] \hspace{-0.6cm}
& %\hspace{-1cm}
${\sigma}$ \hspace{-0.6cm}
&
$A \sigma_{s,{\rm eff}} / (L i_{\rm ref})$
\vspace{0.1cm}
\\
% {Ohmic resistance} [$\Omega$] \hspace{-0.5cm}
{Ohmic drop} [V] \hspace{-0.6cm}
& %\hspace{-1cm}
$r_f$ \hspace{-0.6cm}
&
$R_f i_{\rm ref} / A$
\vspace{0.1cm}
\\
% \ri{Transference number-related} [-] \hspace{-0.5cm}
% {Transf. number-related} [C$\cdot$m$^2\cdot$mol$^{-1}$] \hspace{-0.5cm}
{Transf. number-related [m$^{-2}$]} \hspace{-0.6cm}
& %\hspace{-1cm}
$\gamma$ \hspace{-0.6cm}
&
% $1-t_+$
% $(1-t_+) / (F A)$
{$(1-t_+) / A$}
\vspace{0.1cm}
\\
{Activity coeff.-related} [-] \hspace{-0.6cm}
& %\hspace{-1cm}
$k_{\gamma f}$ \hspace{-0.6cm}
&
$\gamma ( 1+ d \ln f_{\pm} / d \ln c_{e} )$
\vspace{0.1cm}
\\
% Lithium moles time [s] \hspace{-0.6cm}
% & %\hspace{-1cm}
% $\tau_{n_{Li}}$ \hspace{-0.6cm}
% &
% $F n_{Li} / i_{\rm ref}$
% \vspace{0.1cm}
% \\
% \gi{Neg. electrode-separator $\varepsilon_e$ ratio} [-]
% & %\hspace{-1cm}
% $\varepsilon_e^{-s}$
% &
% $\varepsilon_e^- / \varepsilon_e^s$
% \vspace{0.1cm}
% \\
% \gi{Separator-pos. electrode- $\varepsilon_e$ ratio} [-]
% & %\hspace{-1cm}
% $\varepsilon_e^{s+}$
% &
% $\varepsilon_e^- / \varepsilon_e^s$
% \vspace{0.1cm}
% \\
% {Norm. $\delta^-$} [-] \hspace{-0.5cm}
% & %\hspace{-1cm}
% $\bar{\delta}^-$ \hspace{-0.5cm}
% &
% $\delta^- / \ell$
% \vspace{0.1cm}
% \\
% {Norm. $\delta^s$} [-] \hspace{-0.5cm}
% & %\hspace{-1cm}
% $\bar{\delta}^s$ \hspace{-0.5cm}
% &
% $\delta^s / \ell$
% \vspace{0.1cm}
% \\
\multicolumn{3}{c}{Secondary grouped parameters} \\
{Thermal voltage} [V] \hspace{-0.6cm}
& %\hspace{-1cm}
$\beta^{-1}$ \hspace{-0.6cm}
&
$R_g T / (\alpha F)$
\vspace{0.1cm}
\\
{Specific interfacial area} [m$^{-1}$] \hspace{-0.6cm}
& %\hspace{-1cm}
$a_s$ \hspace{-0.6cm}
&
$3 \varepsilon_{s} / R_s$
\vspace{0.1cm}
\\
{Filler volume fraction} [-] \hspace{-0.6cm}
& %\hspace{-1cm}
$\varepsilon_{\rm f}$ \hspace{-0.6cm}
&
$1 - \varepsilon_s - \varepsilon_e$
\vspace{0.1cm}
\\
% {Effective electrolyte diffusion} [m$^2\cdot$s$^{-1}$]
% & %\hspace{-1cm}
% $D_{e,{\rm eff}}$
% &
% $(\varepsilon_e)^b D_e$
% \vspace{0.1cm}
% \\
{Effective property} $\Phi_e \in \{D_e, \kappa_e\}$ \hspace{-0.6cm}
& %\hspace{-1cm}
$\Phi_{e,{\rm eff}}$ \hspace{-0.6cm}
&
$(\varepsilon_e)^b \Phi_e$
\vspace{0.1cm}
\\
{Effective electronic conductivity}  \hspace{-0.6cm}
& %\hspace{-1cm}
$\sigma_{s,{\rm eff}}$ \hspace{-0.6cm}
&
$\varepsilon_s \sigma_s$
\vspace{0.1cm}
\\
{Cum. region thickness N.E.} [m] \hspace{-0.6cm}
& %\hspace{-1cm}
$\delta^-$ \hspace{-0.6cm}
&
$L^-$
\vspace{0.1cm}
\\
{Cum. region thickness N.E.-S.} [m] \hspace{-0.6cm}
& %\hspace{-1cm}
$\delta^s$ \hspace{-0.6cm}
&
$L^- + L^s$
\vspace{0.1cm}
\\
{Cell thickness} [m] \hspace{-0.6cm}
& %\hspace{-1cm}
$\ell$ \hspace{-0.6cm}
&
$L^- + L^s + L^+$
\vspace{0.1cm}
\\
\multicolumn{3}{c}{Normalized variables} 
\vspace{0.1cm}
\\
{Solid-phase concentration} \hspace{-0.6cm}
& %\hspace{-1cm}
$\bar{c}_s$ \hspace{-0.6cm}
&
$c_s / c_{s,{\rm max}}$
\vspace{0.1cm}
\\
{Electrolyte-phase concentration} \hspace{-0.6cm}
& %\hspace{-1cm}
$\bar{c}_e$ \hspace{-0.6cm}
&
$c_e / c_{e,{\rm ref}}$
\vspace{0.1cm}
\\
{Radial dimension} \hspace{-0.6cm}
& %\hspace{-1cm}
$\bar{r}$ \hspace{-0.6cm}
&
$r / R_s$
\vspace{0.1cm}
\\
{Coordinate $x$ } \hspace{-0.6cm}
& %\hspace{-1cm}
$\bar{x}$ \hspace{-0.6cm}
&
% $x / \ell$
$x / L$
\vspace{0.1cm}
\\
{Reaction rate current}  \hspace{-0.6cm}%[A]
& %\hspace{-1cm}
$\tilde{j}_n$ \hspace{-0.6cm}
&
$F a_s L A j_n / i_{\rm ref}$
\vspace{0.1cm}
\\
{Exchange current}  \hspace{-0.6cm}%[A]
& %\hspace{-1cm}
$\tilde{i}_0$ \hspace{-0.6cm}
&
$a_s L A i_0 / i_{\rm ref}$
\vspace{0.1cm}
\\
{Ionic current}  \hspace{-0.6cm}%[A]
& %\hspace{-1cm}
$\tilde{i}_e$ \hspace{-0.6cm}
&
$A i_{e} / i_{\rm ref}$
\vspace{0.1cm}
\\
{Electronic current}  \hspace{-0.6cm}%[A]
& %\hspace{-1cm}
$\tilde{i}_s$ \hspace{-0.6cm}
&
$A i_{s} / i_{\rm ref}$
\vspace{0.1cm}
\\
{Applied current}  \hspace{-0.6cm}%[A]
& %\hspace{-1cm}
$\tilde{i}$ \hspace{-0.6cm}
&
$A i / i_{\rm ref}$
\vspace{0.1cm}
\\
% \multicolumn{3}{c}{Redefined variables \ri{maybe normalize these?}} 
% \vspace{0.1cm}
% \\
% {Reaction rate current} [A]
% & %\hspace{-1cm}
% $\tilde{j}_n^\pm$
% &
% $F a_s^\pm L^\pm A j_n^\pm$
% \vspace{0.1cm}
% \\
% {Exchange current} [A]
% & %\hspace{-1cm}
% $\tilde{i}_0^\pm$
% &
% $a_s^\pm L^\pm A i_0^\pm$
% \vspace{0.1cm}
% \\
% {Ionic current} [A]
% & %\hspace{-1cm}
% $\tilde{i}_e$
% &
% $A i_{e}$
% \vspace{0.1cm}
% \\
% {Electronic current} [A]
% & %\hspace{-1cm}
% $\tilde{i}_s$
% &
% $A i_{s}$
% \vspace{0.1cm}
% \\
% {Applied current} [A]
% & %\hspace{-1cm}
% $\tilde{i}$
% &
% $A i$
% \vspace{0.1cm}
% \\
\hline
\end{tabular}
$^a$Parameter associated to input in electrolyte diffusion equation.

	\label{t:grouped_params}
%\vspace{-0.6cm}
\end{table}
%%------------------------------------------------

Together, these six equations form a comprehensive model that captures the key physical and chemical processes occurring in a lithium-ion battery, making the P2D model an essential tool for battery research. 
Notice how this model is composed of 
%25 parameters 
{30 parameters} in total for solid and electrolyte phase as well as negative electrode, separator and positive electrode regions (see Table \ref{t:phys_params}) 
assuming that:
\begin{enumerate}[label={}, leftmargin=1cm]
\item[($\mathcal{A}$3)] Properties like electrolyte diffusion coefficient and ionic conductivity are independent of space.
\item[($\mathcal{A}$4)] Constants and coefficients are known.
\end{enumerate}

%\vspace{-0.1cm}

\subsection{Single-Particle Model}

%\vspace{-0.1cm}

The SPM is a reduced-order electrochemical model. 
In contrast with the P2D model, the SPM assumes that each electrode in a battery can be represented by a single spherical particle whose surface area is equivalent to the active material surface area of the porous electrode. 
Therefore, the solid-phase diffusion PDE \eqref{spd_new_app} is not a function of ${x}$ but only of ${r}$ and time $t$. 
Moreover, the reaction rate is assumed to be uniform along the $x$-axis, which implies that the pore-wall molar flux $j_n({x},t)$ is obtained by scaling the battery {applied current} ${I}(t)$ as
%\vspace{-0.2cm}
\begin{equation}
    j_n^m (t) \approx \frac{1}{F a_s^m L^m {A}} {I}(t), %i(t),
    \label{jn_spm}
%\vspace{-0.2cm}
\end{equation}
where model parameters are defined in Table \ref{t:phys_params}.

The outlined assumptions require the adaptation of the output voltage equation, which in the case of the SPM takes the form
%\vspace{-0.2cm}
\begin{equation}
v(t) = U_s^+(t) - U_s^-(t) + \eta_s^+(t) - \eta_s^-(t) - \frac{R_f}{{A}} {I}(t),
\label{v_spm}
%\vspace{-0.2cm}
\end{equation}
where the OCP function $U_s^m$ is empirical, and the surface overpotential $\eta_s^m$ can be obtained by solving for this variable the equation of $j_n^m$ in \eqref{kn_new_app} while considering \eqref{jn_spm}, yielding
%\vspace{-0.2cm}
\begin{equation}
    \eta_s^m(t) = \beta^{-1} \sinh^{-1} \left( \frac{1_m}{2 L^m A a_s^m i_{0}^m(t)} {I}(t) \right),
%\vspace{-0.2cm}
\end{equation}
% \begin{equation}
%     i_{0}^\pm = F k^\pm c_{\rm max}^\pm \sqrt{ c_{\rm el} \bar{c}_{ss}^\pm ( 1 - \bar{c}_{ss}^\pm ) }
% \end{equation}
where $1_m = \{-1, +1\}$ for $m = \{+, -\}$, respectively, and the exchange current density $i_0$ is defined in \eqref{kn1_new_app} but without the $x$ dependency.

% %%------------------------------------------------
% \begin{table}[!htb]
% 	\caption{SPM equations$^a$}
% 	\centering
% 	\input{tabs/spm}
% 	\label{t:spm}
% \end{table}
% %%------------------------------------------------

This SPM benefits from being a simple model with fewer parameters 
% (e.g. 14) 
(e.g. {14 parameters}) 
than the P2D model.
Even if the SPM is already a reduced order version of the P2D, it can be further simplified under the assumption that the solid-phase diffusion and the reaction kinetics are arbitrarily fast that holds for very low current rates (a.k.a. pseudo-equilibrium, e.g. below ${\rm C}/20$ current rates). 
Under these conditions, $c_{ss}(t) \approx c_{s,{\rm avg}}(t)$ and by using the PDE in \eqref{spd_new_app} to integrate \eqref{spd2_new_app} while considering the boundary conditions \eqref{spd1_new_app}, the following ordinary differential equation (ODE) is obtained
%\vspace{-0.2cm}
\begin{equation}
    \dot{c}^m_{s,{\rm avg}}(t) = - \frac{1}{F \varepsilon_s^m L^m A} {I}(t).
%\vspace{-0.2cm}
\end{equation}
The associated voltage equation is the equilibrium version of \eqref{v_spm}, i.e.
%\vspace{-0.2cm}
\begin{equation}
v(t) = U_s^+(t) - U_s^-(t).
\label{v_spm_eq}
%\vspace{-0.2cm}
\end{equation}
This version of the SPM is denoted as SPM$_{\rm eq}$.

%\vspace{-0.2cm}

\subsection{Additional Features}

%\vspace{-0.1cm}

Both electrochemical models described above (P2D and SPM) are standard ones as originally presented by different authors. Extensions to these models have been proposed to account for additional effects affecting battery operation, such as thermal dependency, which is relevant in our case as evidenced in \mbox{Fig. \ref{f1}} by the high temperatures reached. 
Another feature that is visible in the data is the CV operations (to charge and discharge the battery), %which is not trivial to capture \gi{\cite{something}} and 
that also needs to be modeled properly. The modeling of these two additional features is now presented.

\subsubsection{Temperature Dependency}

A simple way to account for thermal functionalities is to include temperature-dependent parameters. We resort to an Arrhenius equation of the form:
%\vspace{-0.2cm}
\begin{equation}
    \Phi(t) = \Phi_{\rm ref} \exp \left( \frac{E_\Phi}{R_g} \left( \frac{1}{T_{\rm ref}} - \frac{1}{T(t)} \right) \right),
%\vspace{-0.2cm}
\end{equation}
where $\Phi \in \{ k_n, D_s \}$ and the temperature $T$ corresponds to the measured surface temperature of the cell. 
These thermal dynamics are included in the P2D model, which is denoted as P2DT.

\subsubsection{Constant Voltage}

The CV part of the curve is captured by adding a CV controller in the model during the data sections where a CV charge or discharge steps are applied. This ensures an appropriate CV mode for the model dynamics that slightly differs from that one of the data presumably due to model mismatch. 
%The proportional controller is given by
%\vspace{-0.2cm}
%\begin{equation}
%    {I_{\rm ctrl}}(t) = K_p (v_{\rm ref} - v(t)),
%\vspace{-0.2cm}
%\end{equation}
%where $v_{\rm ref} = \overline{v}$ or $v_{\rm ref} = \underline{v}$ for a maximum voltage $\overline{v}$ CV charge or minimum voltage $\underline{v}$ CV discharge, respectively. Then, $I(t) = {I_{\rm ctrl}}(t)$ during CV. 
The proportional-integral-derivative (PID) controller with $K_p, K_i$ and $K_d$ gains is given by
%\vspace{-0.2cm}
\begin{equation}
    {I_{\rm ctrl}}(t) = K_p e(t) + K_i \int e(\tau) d\tau + K_d \frac{d e(t)}{d t},
%\vspace{-0.2cm}
\end{equation}
where $e(t) = v_{\rm ref} - v(t)$, 
$v_{\rm ref} = \overline{v}$ or $v_{\rm ref} = \underline{v}$ for a maximum voltage $\overline{v}$ CV charge or minimum voltage $\underline{v}$ CV discharge, respectively. Then, $I(t) = {I_{\rm ctrl}}(t)$ during CV. 
This controller is used in both the P2D and SPM.

%\vspace{-0.2cm}

\subsection{Models Normalization}

%\vspace{-0.1cm}

%\ri{Given the inherent interdependencies among certain P2D parameters, isolating individual parameters can be complex. Employing a grouped parameter approach could streamline this process, facilitating more efficient management and analysis. Details of the grouped parameters along with normalized variables, are summarized in Table \ref{t:gp}. The P2D equations, reordered and rewritten according to the grouped parameters, are presented in Table \ref{t:p2d}.}

Both presented P2D and SPM are the as-made models as reported in the literature (see e.g. \cite{doyle1993modeling, fuller1994simulation,santhanagopalan2006online,moura2014adaptive}). 
However, from a quick inspection, it becomes evident that there is a high number of parameters to be identified (see Table \ref{t:phys_params}) either from data or direct experimentation. 
Given the sheer number of parameters, one open question is how identifiable are these parameters, knowing that some of them appear multiplying each other (for instance $1-t_+$ and $a_s$ in \eqref{mb_new_app}). 
Moreover, some parameters are function evaluations which prohibits their analysis in a general setting. For instance, the particle radius $R_s$ in \eqref{spd1_new_app} evaluates the PDE in \eqref{spd_new_app} at the boundary. 
If the radial dimension $r$ is normalized with respect to the radius $R_s$, then the change of variables $\bar{r} = r / R_{s}$ make explicit the PDE \eqref{spd_new_app} in terms of $R_s$ and the characteristic grouped parameter $\tau_s = R_s^2 / D_s$ arises bringing with it another inherent interdependency between the physical parameters $D_s$ and $R_s$.

In order to make parameter relationships evident and facilitate identifiability analysis, we proceeded to normalize the model equations. The normalized P2D model (which is the one used from now on) is reported in Table \ref{t:p2d_norm}, which was obtained using the change of variables and parameter groupings presented in Table \ref{t:grouped_params}. 
The P2D model \cite{doyle1993modeling, fuller1994simulation} in \mbox{Table \ref{t:p2d_app}} 
is completely equivalent (1-to-1 relation) to the normalized version in Table \ref{t:p2d_norm}, but the latter easily shows relationships between parameters. 
{After} this parameter grouping, the P2D model %still 
holds 
%23 \gi{check this} parameters 
{20 parameters (2/3 of the original P2D)} 
while the SPM requires 7 parameters {(1/2 of the original SPM)} as presented in Table \ref{t:num_pars} for convenience.

%\vspace{-0.2cm}

\begin{table}[H]
\caption{Number of model parameters.}
\vspace{-0.5cm}
\label{t:num_pars}
\begin{center}
\begin{tabular}{c c c}
\hline
Model   &As-made   &This work   \\%&\cite{Khalik-2021} \\
\hline
% P2D     &30        &20 \\
% SPM     &14        &7 \\
P2D     &30        &20         \\ %&25 \\
SPM     &14        &7          \\ %&- \\
\hline
\end{tabular}
\end{center}
\end{table}

%\vspace{-0.4cm}

%\begin{remark}
It should be mentioned that there is not a single way of grouping parameters as evidenced in other works \cite{drummond2020structural,Khalik-2021}. 
Several parameterizations were {tested in this work} and the one shown here {required the least amount of parameters.} %for the purpose of this work. 
The SPM helped to define some of the grouped parameters of the P2D model since it has less parameters \cite{bizeray2018identifiability} and the SPM is an asymptotic version of the P2D \cite{Marquis-2019}. 
Moreover, the electrolyte phase is particularly difficult to handle, given that it involves the coupling of three different regions. 
{Indeed}, it was difficult to conciliate the diffusion term of the {in-domain} PDE in \eqref{mb_new_app} with the flux term of the boundary conditions \eqref{mb1_new_app}-\eqref{mb2_new_app} as done for the solid-phase diffusion equations. 
This is because there is not {a straightforward way to modify the boundaries without affecting the in-domain dynamics. 
The only way found so far was to define the input coefficient $\nu_e = F \varepsilon_e L c_{e,{\rm ref}} / i_{\rm ref}$ instead of an electrolyte charge time $\tau_{c,e} = F \varepsilon_e L A c_{e,{\rm ref}} / i_{\rm ref} = A \nu_e$ compatible with $\tau_{c,s}$, and to associate $A$ to the transference number parameter $\gamma = (1-t_+)/A$.} 
%Otherwise, attributing $FA$ to the boundaries would affect the in-domain in an inconsistent fashion.}
%was the most consistent one 
%enough parameters in the boundary to redefine an appropriate set of grouped parameters, and the electrolyte volume fraction appears differently in the PDE and in the boundaries.   
%\end{remark}

%\vspace{-0.1cm}

\section{MODEL IDENTIFICATION}

%\vspace{-0.1cm}

In this section we introduce the methodology exploited in this paper to estimate parameters for the P2D model. In particular, the structural identifiability of the model is analyzed and then a procedure for parameter identification is proposed.

%\vspace{-0.2cm}

\subsection{Structural Identifiability}

%\vspace{-0.1cm}

The identifiability properties of the considered system are assessed through a methodology inspired by structural identifiability as defined in \cite{Ljung-1999}
%,Alavi-2017} \gi{Refrence dont exist} 
as follows:

\begin{definition}[Structural identifiability] \label{def1}
Consider a model structure $\mathcal{M}$ with specific model $\mathcal{M}(t,{\theta})$ parameterized by ${\theta} \subset \mathbb{R}^n$ where $n$ denotes the number of parameters of the model. The identifiability equation for $\mathcal{M}$ is given by:
\vspace{-0.2cm}
\begin{equation}
    \label{Mt}
    \mathcal{M}(t,{\theta}) = \mathcal{M}(t,{\theta}') %\ {\rm for \ almost \ all} \ s
\vspace{-0.1cm}
\end{equation}
% where $\bm{\theta},\bm{\theta}' \in \mathcal{P}$. 
for almost any ${\theta}' \in \mathcal{P}$. 
The model structure $\mathcal{M}$ is said to be
\begin{itemize}
    \item globally identifiable if \eqref{Mt} has a unique solution in $\mathcal{P}$,
    \item locally identifiable if \eqref{Mt} has a finite number of solutions in $\mathcal{P}$,
    \item unidentifiable if \eqref{Mt} has a infinite number of solutions in $\mathcal{P}$.
\end{itemize}
\end{definition}

A structurally globally identifiable model in theory admits a unique solution for the parameter estimation problem and the parameters can be estimated from data. 
%Once structural identifiability is verified for a given number of parameters, then the model parameters can be estimated from data.

To assess the structural identifiability properties of the model, we will resort to the basis of \mbox{Definition \ref{def1}} but applied in a more practical sense. Most approaches used to verify identifiability require the evaluation of the whole model and equations within \cite{Pronzato-1997}. Here, we have identified some characteristic \textit{grouped} parameters that relate to the \textit{physical} parameters, and proceed to show how a specific combination of physical parameters result in exactly the same grouped parameters, which therefore yields exactly the same model response. %that was validated in simulation but not shown here due to space constraints.
%The results were also verified in simulation but are not presented here due to space constraints.

%\vspace{-0.2cm}

\subsection{Parameter Identification Methodology} \label{s:method}

%\vspace{-0.1cm}

Once the model has been normalized to verify its structural identifiability, model parameters might still be too many to handle at the same time. Indeed, a parameter identification problem as an optimization problem of the form \eqref{opt_gen0} may scale poorly with the size of the parameter vector. 
One way to circumvent this difficulty is to leverage the fact that some model dynamics are very little active during certain operating conditions. 
Moreover, different models are more suitable for given operating conditions, which can allow for more efficient parameter estimation. 
These notions are exploited here. 
Four scenarios as a function of the current magnitude can be distinguished, which are reported in Table \ref{t:sce} and explained as follows:
\begin{enumerate}
    \item Under \textit{very low} (pseudo-equilibrium) currents, the battery behaves in a pseudo-equilibrium condition, meaning that concentration and potential gradients are basically uniform and there is no thermal stress. The SPM$_{\rm eq}$ is used with
    % \[{\theta}_{\rm eq} = [q_s^+ \;\; q_s^-]^\top \in \mathbb{R}^2.\]
\vspace{-0.2cm}
    \[{\theta}_{\rm eq} = [\tau_{c,s}^+ \;\; \tau_{c,s}^-]^\top \in \mathbb{R}^2.
\vspace{-0.2cm}\]
    \item \textit{Low} currents allow for relaxation of solid-phase concentration gradients but both electrolyte and thermal dynamics are still assumed to be negligible. 
    The SPM is used with
\vspace{-0.2cm}
    \[{\theta}_{\rm s} = [\tau_{d,s}^+ \;\; \tau_{d,s}^- \;\; \tau_k^+ \;\; \tau_k^- \;\; r_f]^\top \in \mathbb{R}^5.
\vspace{-0.2cm}\]
    \item \textit{Mid} currents form considerable gradients in concentration and potential in solid and electrolyte phases with still limited thermal perturbation. 
    The P2D is used with\footnote{%\mi
    {Only these two parameters are considered here following assumptions $\mathcal{A}_3$ and $\mathcal{A}_4$ and for the sake of simplicity, but this is without loss of generality.}}
\vspace{-0.2cm}
    {\[{\theta}_{\rm e} = [\kappa \;\; \tau_{d,e}]^\top \in \mathbb{R}^2.
\vspace{-0.2cm}\]}   %\;\; \nu_e]^\top \in \mathbb{R}^3.\]}
    \item \textit{High} currents finally excite thermal dynamics.
    The P2DT is used with
\vspace{-0.2cm}
    \[{\theta}_{\rm T} = [E_{\tau_s^+} \;\; E_{\tau_s^-} \;\; E_{\tau_k^+} \;\; E_{\tau_k^-}]^\top \in \mathbb{R}^4.
\vspace{-0.2cm}\]
\end{enumerate}

%\vspace{-0.2cm}

\begin{table}[!htb]
%\vspace{0.3cm}
\caption{Considered scenarios$^\dagger$.}
\vspace{-0.5cm}
\label{t:sce}
\begin{center}
\begin{tabular}{c c c c}
\hline
\multicolumn{3}{c}{Condition} 
    &Model \\
Name    &Abb.   &Ranges 
    & \\
\hline
(Pseudo)-equilibrium &$i_{\rm eq}$ &$i_{\rm eq}(t) \leq i_{{\rm C}/20}$
    &SPM$_{\rm eq}$ \\
Low current &
$i_{\rm s}$ &$i_{{\rm C}/20} < i_{\rm s}(t) \leq i_{{\rm C}/2}$ 
    &SPM \\
Mid current & $i_{\rm e}$ &$i_{{\rm C}/2} < i_{\rm e}(t) \leq i_{1{\rm C}}$ 
    &P2D \\
High current &$i_{\rm T}$ &$i_{\rm T}(t) > i_{1{\rm C}}$ 
    &P2DT \\
\hline
\end{tabular}\\
$^\dagger$The current value for a given C-rate is $i_{{\rm C}}$.
\end{center}
%\vspace{-0.5cm}
\end{table}

The four scenarios allow to first focus on a subset of parameters to be identified for each considered operating condition, and also to exploit the assumed dynamic behaviour of the battery to use a given type of model in order to accelerate the parameter identification process. 
We then divided the optimization problem \eqref{opt_gen0} into subproblems that are simpler and easier to solve. 
The resulting optimization problems were solved via the sequential quadratic programming algorithm %\cite{Nocedal-2006} 
of the \texttt{fmincon} function in Matlab R2021b\textregistered{} using a 
2.10 GHz 16-core laptop with 32 GB of RAM.

\section{RESULTS \& DISCUSSION}

%\vspace{-0.1cm}

In this section, we first analyze the identifiability properties of the physical parameters for the P2D model, and the implications in the grouped parameters. 
%Only geometric parameters such as $R_s$ and $L$ are considered in this contribution. 
Then, we proceed to identify the model parameters considering different operating conditions and types of models.

%\vspace{-0.1cm}

\subsection{Parameter Identifiability Analysis}

%\vspace{-0.1cm}
%\subsubsection{Physical parameters}

% To assess the structural identifiability properties of the model, we resort to \mbox{Definition \ref{def1}}. Most approaches used to verify identifiability require the evaluation of the whole model and equations within \cite{Pronzato} \gi{(This citation doesn't exist in bib)}. Here, we have identified some characteristic grouped parameters that relate to the physical parameters, and proceed to show how a specific combination of physical parameters result in exactly the same grouped parameters, which therefore yields exactly the same model response. %that was validated in simulation but not shown here due to space constraints.
% The results were also verified in simulation but are not presented here due to space constraints.

{The physical parameters that are considered as modifiable are the ones in the group of rows "type (c)" in Table \ref{t:phys_params}. All the other parameters are either known constants, coefficients or internal model variables. Similarly, the grouped parameters that contain these modifiable parameters as independent variables are the main grouped parameters in Table \ref{t:grouped_params}.} Notice that through inspection, it is possible to see that there is a clear cut between parameters associated to solid phase, such as solid-phase diffusion and kinetics, and the ones associated to electrolyte phase, such as electrolyte-phase diffusion and ionic conduction. The analysis will then follow this distinction and verify this assumption.

Let us first focus on solid phase dynamics, which involves the grouped parameters $\tau_s$ and $\tau_k$ that are considered in terms of modifiable physical parameters as
\vspace{-0.15cm}
\begin{equation}
    \tau_s = \frac{R_s^2}{D_s}, \;\; \bar{\tau}_k = \frac{R_s}{k_n},
\vspace{-0.15cm}
\end{equation}
with $\bar{\tau}_k = \sqrt{c_{e,{\rm ref}}} \tau_k$. 
These parameters can be concatenated and since they appear multiplying each other, we take the logarithm in both sides to obtain the following equation
\vspace{-0.15cm}
\begin{equation}
\left[
\begin{array}{c}
     \log( \tau_s ) \\
     \log( \bar{\tau}_k )
\end{array}
\right] = 
\left[
\begin{array}{ccc}
     2 &-1 &0 \\
     1 &0 &-1
\end{array}
\right]
\left[
\begin{array}{c}
     \log( R_s ) \\
     \log( D_s ) \\
     \log( k_n )
\end{array}
\right].
\vspace{-0.15cm}
\end{equation}
This is an undetermined system leading to an infinite number of solutions. Therefore, by considering $R_s$ as an independent variable without loss of generality, we could take $\hat{R}_s = \mu R_s$, $\mu > 0$, and if the other parameters are scaled such as $\hat{D}_s = \mu^2 D_s$ and $\hat{k}_n = \mu k_n$, then
% \begin{eqnarray}
%     \!\!\!\!\!\!\!\!
%     \displaystyle \frac{R_s^2}{D_s} = \frac{(\mu R_s)^2}{\mu^2 D_s}
%     &\!\!\!\!\!\!\implies&\!\!\!\!\!\!
%     {\tau}_s = \hat{\tau}_s 
%     , \\
%     \!\!\!\!\!\!\!\!
%     \displaystyle \frac{R_s}{k_n} = \frac{\mu R_s}{\mu k_n}
%     &\!\!\!\!\!\!\implies&\!\!\!\!\!\! 
%     \bar{\tau}_k = \hat{\bar{\tau}}_k
%     ,
% \end{eqnarray}
\vspace{-0.15cm}
\begin{equation}
    \displaystyle \frac{R_s^2}{D_s} \!\!=\!\! \frac{(\mu R_s)^2}{\mu^2 D_s}
    \!\!\implies\!\!
    {\tau}_s = \hat{\tau}_s 
    , 
    \displaystyle \frac{R_s}{k_n} \!\!=\!\! \frac{\mu R_s}{\mu k_n}
    \!\!\implies\!\!
    \bar{\tau}_k = \hat{\bar{\tau}}_k
    .
\vspace{-0.15cm}
\end{equation}
%This means that the physical parameters $\{R_s, D_s, k_n\}$ are not identifiable independently. 
From the physical perspective, the proposed parameter scaling means that the effect of having bigger particles will not be seen if the kinetics are increased in the same proportion and the diffusion in those particles is squared to that proportion.

\begin{remark}\label{rmk1}
    The physical parameters $\{R_s, D_s, k_n\}$ associated to the solid-phase diffusion and reaction kinetics are not identifiable independently. The same result is obtained with a similar analysis of an SPM type of model.
\end{remark}

Let us now consider the electrolyte phase dynamics, whose grouped parameters can be written in terms of modifiable physical parameters as
\vspace{-0.15cm}
\begin{equation}
    % \bar{q}_s^m = \varepsilon_s^m L^m A, \;\;
    % \bar{q}_e^m = \varepsilon_e^m L^m A, 
    \bar{\tau}_{c,s}^m = \varepsilon_s^m L^m A, \;\;
    {\bar{\nu}_{e}^m = \varepsilon_e^m L^m}, 
\vspace{-0.15cm}
\end{equation}
with $\bar{\tau}_{c,s}^m = i_{\rm ref} \tau_{c,s}^m / (F c_{s,{\rm max}}^m)$ and 
${\bar{\nu}_{e}^m = i_{\rm ref} \nu_{e}^m / c_{e,{\rm ref}}}$ for {solid-phase} {charge time} {and electrolyte input coefficient, respectively}, while the diffusion time constant $\tau_{d,e}$ %and the speed constant $\nu_e$ are left as they are. 
is left as it is. 
The remaining grouped parameters are not directly considered to evaluate possible unidentifiability because they have independent physical parameters that can absorb parametric changes, such as $\{\kappa_{e}, \sigma_{s}, R_f, {t_+, f_{\pm}}\}$ in $\{\kappa, \sigma, r_f, {\gamma, k_{\gamma f}}\}$. 
Notice that most physical parameters appear multiplying each other in the grouped parameters. Therefore, we take the logarithm as before, resulting in the following set of equations:
\vspace{-0.15cm}
\begin{eqnarray}
    % \log (\tau_{d,e}^m) \!\!\!\!&=\!\!\!\!& 2 \log (\ell) - (b-1) \log (\varepsilon_e^m) - \log(D_e)
    % \label{tauem}
    % , \\
    % \log (\nu_e^m) \!\!\!\!&=\!\!\!\!& \log (\ell) - b \log (\varepsilon_e^m) - \log(D_e)
    % \label{nuem}
    % , \\
\hspace{-0.5cm}    \log (\tau_{d,e}^m) \!\!\!\!&=\!\!\!\!& 2 \log (L^m) - (b-1) \log (\varepsilon_e^m) - \log(D_e)
    \label{tauem}
    , \\
\hspace{-0.5cm}    \log (\nu_e^m) \!\!\!\!&=\!\!\!\!& {\log (L^m) + \log (\varepsilon_e^m)}
    \label{nuem}
    , \\
    % \log (\tau_{c,e}^m) \!\!\!\!&=\!\!\!\!& \log (\varepsilon_e^m) + \log (L^m) + \log(A)
    % \label{qem}
    % , \\
    % \log (\tau_{c,s}^m) \!\!\!\!&=\!\!\!\!& \log (\varepsilon_s^m) + \log (L^m) + \log(A),
    % \label{qsm}
\hspace{-0.5cm}    \log (\tau_{c,s}^m) \!\!\!\!&=\!\!\!\!& {\log (L^m) + \log (\varepsilon_s^m) + \log(A),}
    \label{qsm}
\vspace{-0.15cm}
\end{eqnarray}
where there are three equations of the form \eqref{tauem} and \eqref{nuem} for $m \in \{+,-,s\}$ and 
two equations of the form %\eqref{qem} and 
\eqref{qsm} for $m \in \{+,-\}$ 
for a total of {8} equations. 
This system of equations can be written as
\vspace{-0.15cm}
\begin{equation}
    \log ( \theta_{\rm elyte}^{\rm group} ) = M_1 \log( \theta_{\rm elyte}^{\rm phys} )
\vspace{-0.15cm}
\end{equation}
where
% $$\theta_{\rm elyte}^{\rm group} = [ \tau_e^+ \;\; \tau_e^s \;\; \tau_e^- \;\; \nu_e^+ \;\; \nu_e^s \;\; \nu_e^- \;\; \tau_{c,e}^+ \;\; \tau_{c,e}^- \;\; \tau_{c,s}^+ \;\; \tau_{c,s}^-]^\top \in \mathbb{R}^{10},$$
$${\theta_{\rm elyte}^{\rm group} = [ \tau_{d,e}^+ \;\; \tau_{d,e}^s \;\; \tau_{d,e}^- \;\; \nu_e^+ \;\; \nu_e^s \;\; \nu_e^- \;\; \tau_{c,s}^+ \;\; \tau_{c,s}^-]^\top \in \mathbb{R}^{8},}$$
the matrix $M_1$ takes the form
%{\scriptsize
\vspace{-0.15cm}
\begin{equation}
{
M = 
\!\!\!
\left[\!\!
\begin{array}{cccccccccc}
    2 \!\!&0 \!\!&0 \!\!&(1-b)  \!\!\!\!\!\!&0      \!\!\!\!\!\!&0      \!\!&-1 \!\!&0 \!\!&0 \!\!&0 \\
    0 \!\!&2 \!\!&0 \!\!&0      \!\!\!\!\!\!&(1-b)  \!\!\!\!\!\!&0      \!\!&-1 \!\!&0 \!\!&0 \!\!&0 \\
    0 \!\!&0 \!\!&2 \!\!&0      \!\!\!\!\!\!&0      \!\!\!\!\!\!&(1-b)  \!\!&-1 \!\!&0 \!\!&0 \!\!&0 \\
    1 \!\!&0 \!\!&0 \!\!&1     \!\!\!\!\!\!&0      \!\!\!\!\!\!&0      \!\!&0  \!\!&0 \!\!&0 \!\!&0 \\
    0 \!\!&1 \!\!&0 \!\!&0      \!\!\!\!\!\!&1      \!\!\!\!\!\!&0      \!\!&0  \!\!&0 \!\!&0 \!\!&0 \\
    0 \!\!&0 \!\!&1 \!\!&0      \!\!\!\!\!\!&0      \!\!\!\!\!\!&1      \!\!&0 \!\!&0 \!\!&0 \!\!&0 \\
    % 0 \!\!&1 \!\!&0 \!\!&1      \!\!\!\!\!\!&0      \!\!\!\!\!\!&0      \!\!&0  \!\!&0 \!\!&0 \!\!&1 \\
    % 0 \!\!&0 \!\!&1 \!\!&0      \!\!\!\!\!\!&0      \!\!\!\!\!\!&1      \!\!&0  \!\!&0 \!\!&0 \!\!&1 \\
    1 \!\!&0 \!\!&0 \!\!&0      \!\!\!\!\!\!&0      \!\!\!\!\!\!&0      \!\!&0  \!\!&1 \!\!&0 \!\!&1 \\
    0 \!\!&0 \!\!&1 \!\!&0      \!\!\!\!\!\!&0      \!\!\!\!\!\!&0      \!\!&0  \!\!&0 \!\!&1 \!\!&1 
\end{array}
\!\!\right]
}
\vspace{-0.15cm}
\end{equation}
%}
and
% $$\theta_{\rm elyte}^{\rm phys} = [ \ell \;\; L^+ \;\; L^- \;\; \varepsilon_e^+ \;\; \varepsilon_e^- \;\; \varepsilon_e^s \;\; D_e \;\; \varepsilon_s^+ \;\; \varepsilon_s^- \;\; A ]^\top \in \mathbb{R}^{10}.$$
\vspace{-0.15cm}
$${\theta_{\rm elyte}^{\rm phys} = [ L^+ \;\; L^s \;\; L^- \;\; \varepsilon_e^+ \;\; \varepsilon_e^- \;\; \varepsilon_e^s \;\; D_e \;\; \varepsilon_s^+ \;\; \varepsilon_s^- \;\; A ]^\top \in \mathbb{R}^{10}.}
\vspace{-0.15cm}
$$
Other variables like 
% $L^s$ can be obtained as $L^s = \ell - (L^+ + L^-)$ (and thus $\bar{\delta}^-$ and $\bar{\delta}^s$ can be computed), and $\varepsilon_{\rm f} = 1 - (\varepsilon_e + \varepsilon_s)$.
{$\varepsilon_{\rm f}$ can be obtained as $\varepsilon_{\rm f} = 1 - (\varepsilon_e + \varepsilon_s)$.}

It can be shown that ${\rm rank}(M) = 8$, i.e. there are two equations that are linearly dependent. Therefore, the system is undetermined and an infinite number of solutions can be obtained. Similarly as before, let us propose the following scalings for the thickness variables %$\hat{\ell} = \mu_1 \ell$, 
$\hat{L}^m = \mu_1 L^m$, for electrolyte volume fraction $\hat{\varepsilon}_e^m = \varepsilon_e^m / \mu_1$, 
%$m \in \{+,s,-\}$, 
for solid-phase volume fraction $\hat{\varepsilon}_s^m = \varepsilon_s^m / (\mu_1 {\mu_2})$, for electrolyte diffusion coefficient $\hat{D}_e = \mu_1^{b+1} D_e$ and for the area $\hat{A} = {\mu_2} A$ {with $\mu_1, \mu_2 > 0$}. These scalings ensure that $\theta_{\rm elyte}^{\rm group} \implies \hat{\theta}_{\rm elyte}^{\rm group}$.

% We still need to verify the computation of $\bar{\delta}^-$ and $\bar{\delta}^s$. This is given by
% \begin{equation}
%     \hat{\bar{\delta}}^- = \frac{\hat{L}^-}{\hat{\ell}} = \frac{\mu_2 {L}^-}{\mu_1 {\ell}}, \;\;
%     \hat{\bar{\delta}}^s = \frac{\hat{L}^- + \hat{L}^s}{\hat{\ell}} = 1 - \frac{\mu_2 {L}^+}{\mu_1 {\ell}},
% \end{equation}
% with $\hat{L}^s = \hat{\ell} - (\hat{L}^+ + \hat{L}^-) = \mu_1 {\ell} - \mu_2 ({L}^+ + {L}^-)$. 
% Notice that for ${\bar{\delta}}^- = \hat{\bar{\delta}}^-$ to hold, $\mu_1 = \mu_2 = \mu$. Under this condition, 
% $$1 - \frac{\mu_2 L^+}{\mu_1 \ell} = 1 - \frac{L^+}{\ell} = \frac{L^- + L^s}{\ell},$$
% which also guarantees that ${\bar{\delta}}^s = \hat{\bar{\delta}}^s$. 
% Finally, notice that the only parameter that relies on both $\mu_1$ and $\mu_2$ is $A$, which then yields that $\hat{A} = A$.

Let us now focus on the remaining parameters 
% The value of $\varepsilon_{\rm f}$ can accommodate the values of $\varepsilon_e$ and $\varepsilon_s$ such that  $\hat{\varepsilon}_{\rm f} = \mu - \varepsilon_e - \varepsilon_s$. The other parameters 
that need to be re-scaled {like} $\kappa$, $\sigma$ and $r_f$. Consider
\vspace{-0.1cm}
\begin{equation}
    \bar{\kappa} = \frac{A \varepsilon_e^b \kappa_e}{L}, \;\;
    \bar{\sigma} = \frac{A \varepsilon_s \sigma_s}{L}, \;\;
    \bar{r}_f = \frac{R_f}{A}, 
\vspace{-0.1cm}
\end{equation}
where $\bar{\kappa} = i_{\rm ref} \kappa$, 
$\bar{\sigma} = i_{\rm ref} \sigma$ and 
$\bar{r}_f = r_f /  i_{\rm ref}$. 
Notice that in these cases, the choices of $\hat{\kappa}_e = \mu_1^{b+1} {/ \mu_2} \kappa_e$, $\hat{\sigma}_s = \mu_1^2 \sigma_s$ and ${\hat{R}_f = \mu_2 R_f}$ ensures that $\hat{\bar{\kappa}} = \bar{\kappa}$, $\hat{\bar{\sigma}} = \bar{\sigma}$ and $\hat{\bar{r}}_f = \bar{r}_f$.

{The last parameters are $\gamma$ and $k_{\gamma f}$ given by
\vspace{-0.1cm}
\begin{equation}
    \gamma = \frac{1 - t_+}{F A}, \;\;
    k_{\gamma f} = \gamma k_f,
\vspace{-0.1cm}
\end{equation}
with $k_f = 1 + d \ln f_{\pm} / d \ln c_{e}$. 
The scaling $\hat{t}_+ = 1 - (1-t_+) \mu_2$ guarantees that $\hat{\gamma} = \gamma$ whereas $\hat{k}_f = k_f / \mu_2$ ensures that $\hat{k}_{\gamma f} = {k}_{\gamma f}$.}

Physically, having a thicker cell will not render any difference if the volume fraction in both the electrolyte and the solid phase is reduced proportionally {and in that proportion squared, respectively\footnote{{Assuming $\mu_1 = \mu_2 = \mu$.}},} the electrolyte diffusion and ionic conductivity are increased in that proportion to the power of $b+1$ {and $b$, respectively,} the electronic conductivity is increased in that proportion squared, 
{and film resistance is increased by that proportion (with corresponding transference number and activity coefficient)}.

\begin{remark}\label{rmk2}
    The physical parameters 
    % $\{\ell, L^m, \varepsilon_e^m, D_e, \varepsilon_s^m\}$ associated to electrolyte dynamics 
    {$\{L^m, \varepsilon_e^m, D_e, \varepsilon_s^m, A\}$} %associated to electrolyte dynamics 
    are not identifiable independently. 
    % The area $A$, however, is not affected by a possible change of the other parameters, which makes it identifiable in theory, although the model sensitivity to this parameter should be verified.
    This analysis can only be done in the context of higher order models such as the P2D. 
    These identifiability results (as well as the ones of \mbox{Remark \ref{rmk1}}) were also numerically verified in simulation, but they are not presented here due to space constraints.
\end{remark}

For the sake of completeness, let us touch upon the thermal dynamics case. Since the model used here is very simple (temperature-dependent parameters), the physical parameters appear isolated and there is no need for further analysis. If a more advanced thermal model is used (like \cite{Bernardi-1985}), %,Lin-2014}), 
then a similar study as shown here needs to be performed.

{With the previous analysis we have just shown that a given set of physical parameters cannot be uniquely identified from data. Indeed, different combination of physical parameters will render exactly the same characteristic grouped parameters and also exactly the same model response. 
%However, the grouped parameters do not appear multiplying each other and therefore could be estimated. 
%On the other hand, grouped parameters can be identified from data and if specifically devoted measurements are available (for instance, SEM to determine particle radius), the other physical parameters can be inferred from the additional measurements and identified parameters (for instance, solid-phase diffusion coefficient from the diffusion time constant).}
% However, this does not mean that the group parameters are identifiable per se. 
% We now use a similar analysis as above but focusing on group parameters. 
% Consider first the grouped parameters in solid-phase diffusion and reaction kinetics, i.e.
In this sense, the proposed parameter groupings solve the identifiability issues associated to the physical parameters. 
%On the other hand, grouped parameters can be identified from data and if specifically devoted measurements are available (for instance, SEM to determine particle radius), the other physical parameters can be inferred from the additional measurements and identified parameters (for instance, solid-phase diffusion coefficient from the diffusion time constant).}
However, we recognize that this fact does not mean that the grouped parameters are structurally identifiable per se. This verification is more involved, as the P2D model is a system of algebraically coupled PDEs and the theory behind structural identifiability so far has been developed for relatively small nonlinear models of ODEs \cite{Villaverde-2019} to the best of the authors' knowledge, {preventing verification even after discretization}. We have {confirmed} the structural identifiability of a simple {2nd order} SPM for a subset of parameters such as $\tau_{c,s}$, $\tau_{d,s}$ and $\tau_k$ \cite{Couto-2021}. The general evaluation of a P2D model with grouped parameters is left as future work, and we assume that grouped parameters are structurally identifiable in the following.

\subsection{Identification Results}

%\vspace{-0.1cm}

We now focus our attention on the grouped parameters and estimate the parameter values with the methodology presented in Section \ref{s:method}. The root-mean-squared error was used as performance metric for the fit quality, and the optimization time was also computed. The results are shown in Table \ref{t:res1} for the different scenarios presented in Table \ref{t:sce}. The reference values of the P2D model were obtained by evaluating a set of nominal parameters $\theta_{\rm nom}$ coming from specific measurements of physical properties that were carried out in house and cannot be disclosed. 
The response of the model can be seen in the voltage curves of \mbox{Fig. \ref{f2}} where the black curves are the data and the red curves correspond to P2D$(\theta_{\rm nom})$. The parameters $\theta_{\rm nom}$ needed to be slightly adapted in order to generate a reasonable curve accross operating conditions. The data-model error can reach the hundreds of milivolts as it can be verified in the first column of Table \ref{t:res1}, and the discrepancy tends to increase with higher current rates (see Fig. \ref{f2}).

%\vspace{-0.1cm}

\begin{table}[!ht]
\caption{Performance metrics of model identification procedure$^\dagger$.}
\vspace{-0.5cm}
\label{t:res1}
\begin{center}
\begin{tabular}{c c c c c c c c}
\hline
$\mathcal{D}\backslash\mathcal{M}$ 
                        &\multicolumn{1}{c}{$\theta_{\rm nom}$} 
                        &\multicolumn{2}{c}{$\theta_{\rm spm}$} 
                        &\multicolumn{2}{c}{$\theta_{\rm p2d}$}
                        &\multicolumn{2}{c}{$\left. \theta_{\rm p2d} \right\vert_{\theta_0 = \theta_{\rm spm}^{\rm opt}}$} \\
                        &$e_v$ &$t_{\rm opt}$ &$e_v$ &$t_{\rm opt}$ &$e_v$ &$t_{\rm opt}$ &$e_v$ \\
\hline
%Pseudo-equilibrium	
$i_{\rm eq}$ &%71.72/ 
%(49.92)
91.17 &0.01 &69.01 
%(47.78) 
% &26.21 &36.55 &\ri{15.26} &\ri{34.61} \\
% &\bi{42.29 N} &\bi{36.41 N} &\ri{15.26} &\ri{34.61} \\
% &26.21 &36.55 &\bi{42.29 N} &\bi{36.41 N} \\
&{25.11} &{36.55} &{10.75} &{36.55} \\
% Low current
$i_{\rm s}$ &
146.79  &0.02 &63.93 
&11.12 %16.60
&25.69 %25.24
&5.44 &25.64 \\
% Mid current
$i_{\rm e}$ &
178.85  &- &- &0.76 &30.22 &-&- \\
% High current
$i_{\rm T}$ &
155.01  &- &- &0.59 &25.32 &-&- \\
\hline
\end{tabular}
\\
$^\dagger$$e_v$ is error [mV] and $t_{\rm opt}$ is optimization time [h].
\end{center}
%\vspace{-0.7cm}
\end{table}

\begin{figure}[!htb]
	\centering
	\includegraphics[scale=0.40,trim={4.5cm 1.7cm 1.6cm 2cm},clip]{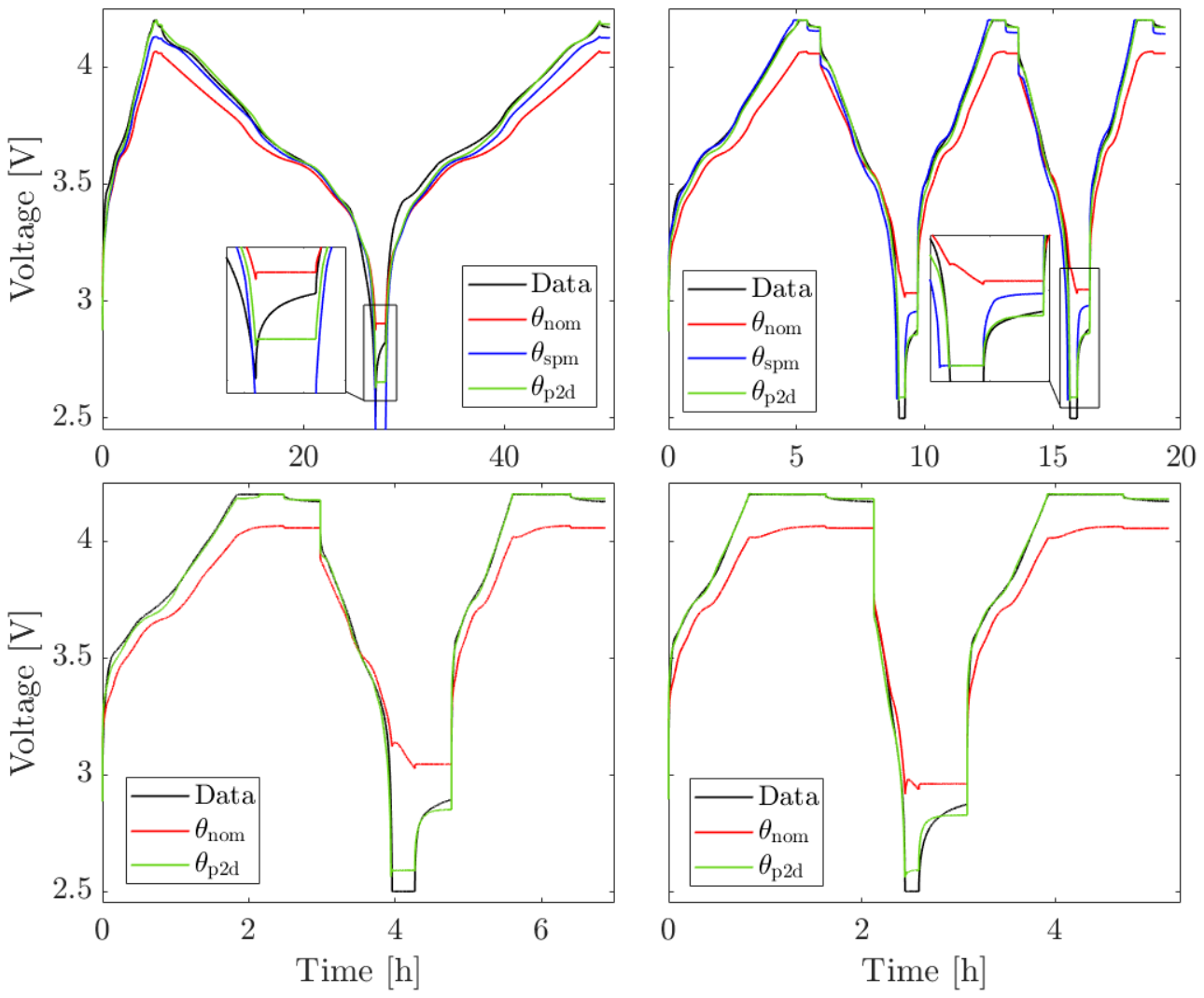}
   \put (-238,200) {a)}
   \put (-121,200) {b)}
   \put (-238,100) {c)}
   \put (-121,100) {d)}
	\vspace{-0.3cm}
	\caption{Voltage profiles from data and model simulations for different optimal parameters under a) ${\rm C}/20$, b) $\{ {\rm C}/5, {\rm C}/3, {\rm C}/2 \}$ (with zoom provided for critical transitions), c) $1{\rm C}$ and d) $3{\rm C}$ currents.}
	\label{f2}
\vspace{-0.3cm}
\end{figure}

Under low current conditions, the SPM is a valid model. Therefore, we used it in order to obtain the model parameters $\theta_{\rm spm}$ that we then used to evaluate the P2D model as P2D$(\theta_{\rm spm})$. This optimization problem was solved in 36 s for the pseudo-equilibrium condition and twice that time for the low currents. The error levels obtained were between 60 mV and 70 mV as shown in Table \ref{t:res1}. To have a sense of the quality of the fit, Fig. \ref{f2}a and \ref{f2}b shows the response obtained from the P2D. Notice how close this response is with respect to the data. This result is benchmarked by comparing it with the result obtained by optimizing the P2D model itself (i.e. P2D($\theta_{\rm p2d}$)), which is also reported in Table \ref{t:res1}. Notice that for the low current scenarios, which are the cases that require the most amount of data (longest experiments), the optimization problems considering P2D$(\theta_{\rm p2d})$ require dozens of hours (500 times more time) but they incur in half of the error compared to the P2D$(\theta_{\rm spm})$. This time difference is considerable. For mid to high currents, the optimization time is more reasonable by being under the hour mark and exhibiting a similar order of magnitude for the error compared to the low current cases. Therefore, leveraging low current conditions and exploiting model assumptions, we can accelerate the estimation of model parameters for the P2D by a 500 factor at least if the doubling of the error can be tolerated by the given application.

\setcounter{table}{0}
\renewcommand{\thetable}{A\arabic{table}}
\setcounter{equation}{0}
\renewcommand{\theequation}{A\arabic{equation}}

%%------------------------------------------------
\begin{table*}[!htb]
 \vspace{0.2cm}
	\caption{P2D model equations with physical parameters defined in Table \ref{t:phys_params}$^\dagger$.}
 \vspace{-0.3cm}
	\centering
	
{\footnotesize
%{\small
\begin{tabular}{l c c c}
\hline
\textbf{Name} \hspace{-0.7cm} & \textbf{Equation} &%\hspace{2cm} 
\textbf{Boundary conditions/additional equations} &
\hspace{-1cm} \textbf{Eq.}\\ 
\hline
\begin{minipage}{1.5cm}
Solid-phase\\
diffusion
\end{minipage}
 \hspace{-0.5cm}
&
\parbox{7cm}{
\begin{equation} 
% \displaystyle \frac{\partial {c}_{s}^\pm}{\partial t} = 
% \frac{1}{r^2} \frac{\partial}{\partial r} 
% \left( r^2 D_s \frac{\partial c_s}{\partial r} \right)
\displaystyle \frac{\partial {c}_{s}}{\partial t} ({x},{r},t) = 
%\frac{1}{\tau_{d,s}^m} 
% \left( \frac{2}{\bar{r}} \frac{\partial \bar{c}_{s}^\pm}{\partial \bar{r}}(\bar{x},\bar{r},t) + 
% \frac{\partial^2 \bar{c}_{s}^\pm}{\partial \bar{r}^2}(\bar{x},\bar{r},t) \right)
\frac{1}{{r}^2} \frac{\partial}{\partial {r}} \left( \! {r}^2 D_s^m \frac{\partial {c}_s^m}{\partial {r}} ({x},{r},t) \! \right)
\nonumber
\end{equation}
}
& \hspace{-0.2cm}
\parbox{8.2cm}{
% \begin{eqnarray} 
\begin{subequations}
\label{spd_new_app}
\begin{gather}
% \displaystyle \left. \frac{\partial \bar{c}_s^\pm}{\partial \bar{r}} \right\vert_{\bar{r}=0} &\!\!\!\!\!\!\!\! = &\!\!\!\! 0, \;
\displaystyle 
\left. \frac{\partial {c}_s^m}{\partial {r}}({x},{r},t) \right\vert_{{r}=0} \!\!\!\!\!\!\!\!= 0, \;
\displaystyle  
 \left. \frac{\partial {c}_s^m}{\partial {r}}({x},{r},t) \right\vert_{{r}=R_s^m} \!\!\!\!\!\!\!\!= -\frac{1}{ %q_s^m
 D_s^m} {j}_n^m({x},t) 
 % \nonumber
\label{spd1_new_app}
  \\
 %\bar{c}_{s,{\rm avg}}^m ({x},t) = \displaystyle -\frac{1}{q_s^m} i(t) 
 {c}_{s,{\rm avg}}^m ({x},t) = \displaystyle \frac{3}{(R_s^m)^3} \int_0^{R_s^m} {r}^2 {c}_{s} ({x},r,t) {\rm d} {r} 
 % \nonumber
\label{spd2_new_app}
 \\
 {c}_{ss}^m ({x},t) = \left. {c}_{s}^m ({x},{r},t) \right\vert_{{r}=R_s^m}
 % \nonumber
\label{spd3_new_app}
% \label{spd}
%\nonumber
% \\
% \displaystyle  
%  \left. \frac{\partial \bar{c}_s^\pm}{\partial \bar{r}} \right\vert_{\bar{r}=1} &\!\!\!\!\!\!\!\!= &\!\!\!\! -\frac{\tau_{s}^\pm}{3 q^\pm} J_n^\pm
% \nonumber
\end{gather}
\end{subequations}
% \end{eqnarray}
}
% & %\hspace{1.5cm}
% \parbox{0.5cm}{
% \begin{eqnarray} 
% \hspace{-0.3cm}
% \label{spd_new}
% \end{eqnarray}
% }
\vspace{-0.3cm}
\\
\begin{minipage}{1.5cm}
Kinetics%$^a$
\end{minipage}
 \hspace{-0.7cm}
&
\parbox{7cm}{
\begin{eqnarray} 
%    \tilde{j}_n^m(\bar{x}^m,t) &\!\!\!\!=&\!\!\!\! \tilde{i}_0^m (\bar{x}^m,t) \left( \exp \left( \beta \eta^m (\bar{x}^m,t) \right) - \exp \left( -\beta \eta^m (\bar{x}^m,t) \right) \right)
    {j}_n^m({x},t) &\!\!\!\!=&\!\!\!\! {i}_0^m ({x},t) \left( e^{ \displaystyle \beta \eta^m ({x},t) } - e^{ \displaystyle -\beta \eta^m ({x},t) } \right)
    \nonumber
    % \\
    % I_0^\pm &\!\!\!\!=&\!\!\!\! 3 \tau_{k}^\pm q_{s}^\pm \sqrt{\bar{c}_e \bar{c}_{s}^\pm (1-\bar{c}_{s}^\pm)}
    % \nonumber
    % \\
    % \eta_s^\pm &\!\!\!\!=&\!\!\!\! \phi_{s}^\pm - \phi_{e} - U_s^\pm 
    % \nonumber
\end{eqnarray}
}
& \hspace{-0.2cm}
\parbox{8.2cm}{
% \begin{eqnarray} 
\begin{subequations}
\label{kn_new_app}
\begin{gather}
    {i}_0^m ({x},t) \!=\! %3 %q_{s}^m
%    \tau_{c,s}^m 
	F k_n^m \!
	\left( {c}_e^m ({x},t) \right)^\alpha \!\! \left( {c}_{ss}^m ({x},t) \right)^{(1-\alpha)} \!\! \left( c_{s,{\rm max}}^m \!-\! {c}_{ss}^m ({x},t) \right)^\alpha  %/ \tau_{k}^m
    % \nonumber
    \label{kn1_new_app}
    \\
    \eta^m({x},t) = \phi_{s}^m ({x},t) - \phi_{e}^m ({x},t) - U^m ({x},t) 
    % \nonumber
    \label{kn2_new_app}
% \end{eqnarray}
\end{gather}
\end{subequations}
}
% & %\hspace{1.5cm}
% \parbox{0.5cm}{
% \begin{eqnarray} 
% \hspace{-0.3cm}
% \label{kn_new}
% \end{eqnarray}
% }
\vspace{-0.3cm}
\\
\begin{minipage}{1.5cm}
Mass\\
balance %$^b$
\end{minipage}
 \hspace{-0.7cm}
&
\parbox{7cm}{
\begin{equation} 
    \displaystyle \varepsilon_e^{m} \frac{\partial {c}_{e}^{m}}{\partial t} ({x},t) = \frac{\partial}{\partial x} \left( D_{e,{\rm eff}}^{m}  \frac{\partial {c}_{e}^{m}}{\partial {x}^m} ({x},t) \right) + a_s^m (1-t_+) {j}_n^{m} ({x},t)
    \nonumber
\end{equation}
}
& \hspace{-0.2cm}
\parbox{8.2cm}{
% \begin{eqnarray} 
\begin{subequations}
\label{mb_new_app}
\begin{gather}
\displaystyle %\frac{\varepsilon_e^-}
% \frac{1}{\tau_{e}^-}
D_{e,{\rm eff}}^- \left. \frac{\partial {c}_{e}^-}{\partial {x}} ({x},t) \right\vert_{{x}=0} \!\!\!\!\!\!\!\!=
\displaystyle %\frac{\varepsilon_e^+}
% \frac{1}{\tau_{e}^+} 
D_{e,{\rm eff}}^+ \left. \frac{\partial {c}_{e}^+}{\partial {x}} ({x},t) \right\vert_{{x}=\ell} \!\!\!\!\!\!\!\!\!\!= 0, \;\;
% \displaystyle %\frac{\varepsilon_e^-}{\tau_{e}^-}
% % \frac{\varepsilon_e^{-s}}{\tau_{e}^-} 
% \frac{\mi{\nu_e^-}}{\mi{\tau_{d,e}^-}} \left. \frac{\partial \bar{c}_{e}^-}{\partial \bar{x}} (\bar{x},t) \right\vert_{\bar{x} = \bar{\delta}^-} \!\!\!\!\!\!\!\!=...
% \nonumber
\label{mb1_new_app}
\\
\displaystyle 
D_{e,{\rm eff}}^- \left. \frac{\partial {c}_{e}^-}{\partial {x}} ({x},t) \right\vert_{{x} = \delta^-} 
\!\!\!\!\!\!\!\!=
D_{e,{\rm eff}}^s \left. \frac{\partial {c}_{e}^s}{\partial {x}} ({x},t) \right\vert_{{x} = \delta^-}, \;\;
% \nonumber
\label{mb12_new_app}
\\
% \displaystyle \frac{1}{\tau_{e}^-} \left. \frac{\partial \bar{c}_{e}^-}{\partial \bar{x}^-} \right\vert_{\bar{x}^- = 1} &\!\!\!\!\!\!\!\!\!\!\!\!\!\!= &\!\!\!\!\!\! 
%  \frac{1}{\tau_{e}^s} \left. \frac{\partial \bar{c}_{e}^s}{\partial \bar{x}^s} \right\vert_{\bar{x}^s = 0}
% \nonumber 
% \\
% \displaystyle ...= %\frac{\varepsilon_e^s}
% % \frac{1}{\tau_{e}^s}
% \frac{\mi{\nu_e^s}}{\mi{\tau_{d,e}^s}} \left. \frac{\partial \bar{c}_{e}^s}{\partial \bar{x}} (\bar{x},t) \right\vert_{\bar{x} = \bar{\delta}^-}, \;\;
%\frac{\varepsilon_e^s}
% \frac{\varepsilon_e^{s+}}{\tau_{e}^s}
D_{e,{\rm eff}}^s \left. \frac{\partial {c}_{e}^s}{\partial {x}} ({x},t) \right\vert_{{x} = \delta^s} \!\!\!\!\!\!\!\! =
 %\frac{\varepsilon_e^+}
 % \frac{1}{\tau_{e}^+}
D_{e,{\rm eff}}^+ \left. \frac{\partial {c}_{e}^+}{\partial {x}} ({x},t) \right\vert_{{x} = \delta^s}
\label{mb2_new_app}
% \nonumber
% \\
% \displaystyle \frac{1}{\tau_{e}^+} \left. \frac{\partial \bar{c}_{e}^+}{\partial \bar{x}^+} \right\vert_{\bar{x}^+=1} &\!\!\!\!\!\!\!\!\!\!\!\!\!\!= &\!\!\!\!\!\! 
% 0
% \nonumber
% \end{eqnarray}
\end{gather}
\end{subequations}
}
% & %\hspace{1.5cm}
% \parbox{0.5cm}{
% \begin{eqnarray} 
% \hspace{-0.3cm}
% \label{mb_new}
% \end{eqnarray}
% }
\vspace{-0.3cm}
\\
\begin{minipage}{1.5cm}
Potential in \\
solution
\end{minipage}
 \hspace{-0.7cm}
&
\parbox{7cm}{
\begin{equation} 
   \frac{ {i}_e^m } {{\kappa}_{e,{\rm eff}}^{m}} ({x},t) = - \frac{\partial \phi_{e}^{m} }{\partial {x}} ({x},t) + \frac{2 R T }{F} (1-t_+\!) \!\! \left(\! 1\!\!+\!\! \frac{{\rm d} \! \ln f_{\pm}}{{\rm d} \! \ln c_{e}^m} \!\right)\! \frac{\partial \ln c_{e}^{m} }{\partial {x}} ({x},t)
    \nonumber
\end{equation}
}
& \hspace{-0.2cm}
\parbox{8.2cm}{
% \begin{eqnarray} 
\begin{subequations}
\label{psolution_new_app}
\begin{gather}
\displaystyle \left. {i}_{e}^-({x},t) \right\vert_{{x}=0} 
= 
\displaystyle \left. {i}_{e}^+({x},t) \right\vert_{{x}=\ell}
= 
0, 
% \nonumber
\label{psolution1_new_app}
\vspace{-0.3cm}
\\
\displaystyle \left. {i}_{e}^-({x},t) \right\vert_{{x}=\delta^-} 
= 
\displaystyle \left. {i}_{e}^s({x},t) \right\vert_{{x} = \delta^s}
 =
{i}(t), 
% \nonumber
\label{psolution2_new_app}
% \\
% \displaystyle \left. I_{e}^s \right\vert_{\bar{x}^s = 1} &\!\!\!\!\!\!\!\!= &\!\!\!\! \bar{I}, 
% \nonumber
% \\
% \displaystyle \left. I_{e}^+ \right\vert_{\bar{x}^+=1} &\!\!\!\!\!\!\!\!= &\!\!\!\! 0, 
% \nonumber
% \end{eqnarray}
\end{gather}
\end{subequations}
}
% & %\hspace{1.5cm}
% \parbox{0.5cm}{
% \begin{eqnarray} 
% \hspace{-0.3cm}
% \label{psolution_new}
% \end{eqnarray}
% }
\vspace{-0.3cm}
\\
\begin{minipage}{1.5cm}
Potential in\\
solid
\end{minipage}
 \hspace{-0.7cm}
&
\parbox{7cm}{
\begin{eqnarray} 
    {i}_{s}^{m} ({x},t) &\!\!\!\!\!\!=&\!\!\!\!\!\! -{\sigma}^{m}_{s,{\rm eff}} \frac{\partial \phi_{s}^{m}}{\partial {x}} ({x},t), \;
    a_s^m {j}_n^{m}({x},t) = \frac{1}{F} \frac{\partial {i}_{e}^{m}}{\partial {x}} ({x},t), 
    \nonumber 
    \\
    {i}(t) &\!\!\!\!\!\!=&\!\!\!\!\!\! {i}_{s}^{m}({x},t) + {i}_{e}^{m}({x},t)
    \nonumber 
    % \\
    % \bar{I} &\!\!\!\!=&\!\!\!\! I_{s} + I_{e}
    % \nonumber
    % \\
    % J_n &\!\!\!\!=&\!\!\!\! \frac{\partial I_{e}}{\partial \bar{x}}
    % \nonumber
\end{eqnarray} 
}
& \hspace{-0.2cm}
\parbox{8.2cm}{
% \begin{eqnarray} 
\begin{subequations}
\label{psolid_new_app}
\begin{gather}
\displaystyle \left. {i}_{s}^- ({x},t) \right\vert_{{x}=0} 
=
\displaystyle \left. {i}_{s}^+ ({x},t) \right\vert_{{x}=\ell}
=
{i}(t), 
% \nonumber
\label{psolid1_new_app}
\\
\displaystyle \left. {i}_{s}^- ({x},t) \right\vert_{{x}=\delta^-} 
=
\displaystyle \left. {i}_{s}^+ ({x},t) \right\vert_{{x}=\delta^s} 
=
0, 
\label{psolid2_new_app}
% \nonumber
% \\
% \displaystyle \left. I_{s} \right\vert_{\bar{x}^s=1} &\!\!\!\!\!\!\!\!= &\!\!\!\! 0, 
% \nonumber
% \\
% \displaystyle \left. I_{s} \right\vert_{\bar{x}^+=1} &\!\!\!\!\!\!\!\!= &\!\!\!\! \bar{I}, 
% \nonumber
% \end{eqnarray}
\end{gather}
\end{subequations}
}
% & %\hspace{1.5cm}
% \parbox{0.5cm}{
% \begin{eqnarray} 
% \hspace{-0.3cm}
% \label{psolid_new}
% \end{eqnarray}
% }
\vspace{-0.3cm}
\\
% \begin{minipage}{1.5cm}
% Lithium moles
% \end{minipage}
% &
% \parbox{7cm}{
% \begin{equation} 
%     \ri{\tau_{n_{Li}}(t)} = \tau_{c,s}^+ \bar{c}_{s,{\rm avg}}^+(t) + \tau_{c,s}^- \bar{c}_{s,{\rm avg}}^-(t) 
%     \nonumber
% \end{equation}
% }
% & %\hspace{1.5cm}
% \parbox{8cm}{
% % \begin{equation} 
% \begin{subequations}
% \label{nli_new}
% \begin{gather}
% -
% % \nonumber
% \label{nli1_new}
% % \end{equation}
% \end{gather}
% \end{subequations}
% }
% %& %\hspace{1.5cm}
% % \parbox{0.5cm}{
% % \begin{eqnarray} 
% % \hspace{-0.3cm}
% % \label{volt}
% % \end{eqnarray}
% % }
% \vspace{-0.3cm}
% \\
\begin{minipage}{1.5cm}
Voltage
\end{minipage}
 \hspace{-0.7cm}
&
\parbox{7cm}{
\begin{equation} 
    v(t) = \left. \phi_{s}^+ ({x},t) \right\vert_{x=\ell} - \left. \phi_{s}^- ({x},t) \right\vert_{x=0}  - R_{f} {i}(t)
    \nonumber
\end{equation}
}
& \hspace{-0.2cm}
\parbox{8.2cm}{
% \begin{equation} 
\begin{subequations}
\label{volt_new_app}
\begin{gather}
-
\label{volt1_new_app}
% \nonumber
% \end{equation}
\end{gather}
\end{subequations}
}
% & %\hspace{1.5cm}
% \parbox{0.5cm}{
% \begin{eqnarray} 
% \hspace{-0.3cm}
% \label{volt_new}
% \end{eqnarray}
% }
\vspace{-0.1cm}
\\
\hline
\end{tabular}
\\
%$^a$Assuming $\alpha = 1 - \alpha = 1/2$.\\
% $^b$Normalized variables $\bar{\delta}^- = \delta^- / \ell$ and $\bar{\delta}^s = \delta^s / \ell$.\\
$^\dagger$Superscript $m\in\{+,-,s\}$ for positive electrode, negative electrode and separator, respectively.\\
} 
% \\
% \begin{tabular}{l}
% %{\scriptsize $^a$Assuming $\alpha = 1/2$.} \\
% \end{tabular}

	\label{t:p2d_app}
%\vspace{-0.6cm}
\end{table*}
%%------------------------------------------------

One last experiment consisted in using the optimal parameters $\theta_{\rm spm}^{\rm opt}$ obtained using the SPM as initial parameters $\theta_0$ of the optimization problem using the P2D model as in P2D($\left. \theta_{\rm p2d} \right \vert_{\theta_0 = \theta_{\rm spm}^{\rm opt}}$). 
The results are shown in the two last columns of Table \ref{t:res1}. {Notice that by using a good initial seed, the optimization problems for low currents are able to converge faster (c.a. in half of the time) while still keeping an error level as the original problem, which means that if higher accuracy are required, the combined use of reduced- and high-order models can be beneficial for accurate and fast parameter identification.}

 %%------------------------------------------------
 \begin{table*}[!htb]
 \vspace{0.2cm}
 	\caption{Physical variables associated to the as-made P2D model \cite{doyle1993modeling,fuller1994simulation} in Table \ref{t:p2d_app}.}
     \vspace{-0.3cm}
 	\centering

\begin{tabular}{c l c l c l c}
\hline
Type &Definition \hspace{-0.5cm} & Symbol &Definition \hspace{-0.5cm} & Symbol &Definition \hspace{-0.5cm} & Symbol \\ 
\hline
%\multicolumn{6}{c}{Constants} \\
\multirow{1}{*}{(a)}
    &{Faraday's} [C$\cdot$mol$^{-1}$] \hspace{-0.5cm}
& %\hspace{-1cm}
$F$
%\vspace{0.1cm}
%\\
&
{Universal gas} [J$\cdot$mol$^{-1}\cdot$K$^{-1}$] \hspace{-0.5cm}
& %\hspace{-1cm}
$R_g$
&&
\vspace{0.1cm}
\\
\hdashline
\multirow{1}{*}{(b)}
    &{Apparent transfer coefficient} [-] \hspace{-0.5cm}
& %\hspace{-1cm}
$\alpha$
% \vspace{0.1cm}
% \\
% {Specific interfacial area} [m$^{-1}$]
% & %\hspace{-1cm}
% $a_s$
% &
% x
% \vspace{0.1cm}
% \\
&
{Bruggeman coefficient} [-] \hspace{-0.5cm}
& %\hspace{-1cm}
$b$
%\vspace{0.1cm}
%\\
&&
\vspace{0.1cm}
\\
\hdashline
%\multicolumn{6}{c}{Parameters} \\
\multirow{5}{*}{(c)}
    &{Diffusion coefficient} [m$^2\cdot$s$^{-1}$] \hspace{-0.5cm}
& %\hspace{-1cm}
$D$
%\vspace{0.1cm}
%\\
&
{Particle radius} [m$^2$] \hspace{-0.5cm}
& %\hspace{-1cm}
$R$
% \vspace{0.1cm}
% \\
&
{Volume fraction} [-] \hspace{-0.5cm}
& %\hspace{-1cm}
$\varepsilon$
\vspace{0.1cm}
\\
%&
    &{Region thickness} [m] \hspace{-0.5cm}
& %\hspace{-1cm}
$L$
%\vspace{0.1cm}
%\\
&
% {Cell thickness} [m]
% & %\hspace{-1cm}
% $\ell$
% \vspace{0.1cm}
% \\
{Cross-sectional area} [m$^2$] \hspace{-0.5cm}
& %\hspace{-1cm}
$A$
% \vspace{0.1cm}
% \\
&
{Reaction rate} [m$^{2.5}\cdot$mol$^{-0.5}\cdot$s$^{-1}$] \hspace{-0.5cm}
& %\hspace{-1cm}
$k_n$
\vspace{0.1cm}
\\
    &{Transference number} [-] \hspace{-0.5cm}
& %\hspace{-1cm}
$t_+$
%\vspace{0.1cm}
%\\
&
{Activity coefficient} [-] \hspace{-0.5cm}
& %\hspace{-1cm}
$f_\pm$
&
{Ionic conductivity} [S$\cdot$m$^{-1}$] \hspace{-0.5cm}
& %\hspace{-1cm}
$\kappa_e$
% \vspace{0.1cm}
% \\
\vspace{0.1cm}
\\
    & {Electronic conductivity} [S$\cdot$m$^{-1}$] \hspace{-0.5cm}
& %\hspace{-1cm}
$\sigma_s$
% \vspace{0.1cm}
% \\
&
{Ohmic resistance} [$\Omega\cdot$m$^2$] \hspace{-0.5cm}
& %\hspace{-1cm}
$R_f$
% \vspace{0.1cm}
% \\
&
{Activation energy} [J$\cdot$mol$^{-1}$] \hspace{-0.5cm}
& %\hspace{-1cm}
$E$
\vspace{0.1cm}
\\
\hdashline
% \multicolumn{6}{c}{Variables} 
% \vspace{0.1cm}
% \\
\multirow{4}{*}{(d)}
    &{Concentration} [mol$\cdot$m$^{-3}$] \hspace{-0.5cm}
& %\hspace{-1cm}
${c}$
% \vspace{0.1cm}
% \\
&
{Pore-wall molar flux} [mol$\cdot$m$^{-2}\cdot$s$^{-1}$] \hspace{-0.5cm}
& %\hspace{-1cm}
${j}_n$
%\vspace{0.1cm}
%\\
&
{Exchange current density} [A$\cdot$m$^{-2}$] \hspace{-0.5cm}
& %\hspace{-1cm}
${i}_0$
\vspace{0.1cm}
\\
%&
    &{Overpotential} [V] \hspace{-0.5cm}
& %\hspace{-1cm}
$\eta$
%\vspace{0.1cm}
%\\
&
{Open circuit potential} [V] \hspace{-0.5cm}
& %\hspace{-1cm}
$U$
% \vspace{0.1cm}
% \\
&
{Electric potential} [V] \hspace{-0.5cm}
& %\hspace{-1cm}
$\phi$
\vspace{0.1cm}
\\
    &{Ionic current} {density} [A$\cdot$m$^{-2}$] \hspace{-0.5cm}
& %\hspace{-1cm}
$i_e$
% \vspace{0.1cm}
% \\
&
{Electronic current} {density} [A$\cdot$m$^{-2}$] \hspace{-0.5cm}
& %\hspace{-1cm}
$i_s$
%\vspace{0.1cm}
%\\
&
{Applied current} {density} [A$\cdot$m$^{-2}$] \hspace{-0.5cm}
& %\hspace{-1cm}
$i$
\vspace{0.1cm}
\\
%&
    &{Applied current} [A] \hspace{-0.5cm}
& %\hspace{-1cm}
$I$
&
{Temperature} [K] \hspace{-0.5cm}
&
$T$
&\hspace{-0.5cm}&
\vspace{0.1cm}
\\
% \multicolumn{6}{c}{Superscripts}
% \vspace{0.1cm}
% \\
\hdashline
\multirow{1}{*}{(e)}
    &{Positive electrode} \hspace{-0.5cm}
& %\hspace{-1cm}
$+$
% \vspace{0.1cm}
% \\
&
{Negative electrode} \hspace{-0.5cm}
& %\hspace{-1cm}
$-$
% \vspace{0.1cm}
% \\
&
{Separator} \hspace{-0.5cm}
& %\hspace{-1cm}
$s$
\vspace{0.1cm}
\\
% %&
%     &{Nominal}
% & %\hspace{-1cm}
% $0$
% &&
% &&
% \vspace{0.1cm}
% \\
% \multicolumn{6}{c}{Subscripts}
% \vspace{0.1cm}
% \\
\hdashline
\multirow{3}{*}{(f)}
    &{Solid phase} \hspace{-0.5cm}
& %\hspace{-1cm}
$s$
% \vspace{0.1cm}
% \\
&
{Electrolyte phase} \hspace{-0.5cm}
& %\hspace{-1cm}
$e$
% \vspace{0.1cm}
% \\
&
{Filler} \hspace{-0.5cm}
& %\hspace{-1cm}
${\rm f}$
\vspace{0.1cm}
\\
%&
    &{Surface concentration} \hspace{-0.5cm}
& %\hspace{-1cm}
$ss$
%\vspace{0.1cm}
%\\
&
{Average concentration} \hspace{-0.5cm}
& %\hspace{-1cm}
${\rm avg}$
% \vspace{0.1cm}
% \\
&
{Theoretical maximum} \hspace{-0.5cm}
& %\hspace{-1cm}
${\rm max}$
\vspace{0.1cm}
\\
    &{Effective property} \hspace{-0.5cm}
& %\hspace{-1cm}
${\rm eff}$
% \vspace{0.1cm}
% \\
&
{Reference variable} \hspace{-0.5cm}
& %\hspace{-1cm}
${\rm ref}$
&\hspace{-0.5cm}&
\vspace{0.1cm}
\\
\hline
\end{tabular}
\\
$^\dagger$(a) Constants, (b) coefficients, (c) modifiable parameters, (d) variables, (e) superscripts, (f) subscripts.\\

 	\label{t:phys_params}
%\vspace{-0.6cm}
 \end{table*}
 %%------------------------------------------------

%\vspace{-0.15cm}

\section{CONCLUSIONS}

%\vspace{-0.15cm}

A parameter identification scheme aiming at estimating model parameters of a P2D model has been developed. This scheme relies in three key aspects. 
First, the data used for parameter identification was considered in order to include appropriate model components that allowed to capture the data in its wider extent (for instance, CV operation and thermal dynamics). 
Secondly, the model parameters were evaluated and possible structural identifiability conflicts associated to the physical parameters were handled. 
Finally, the different operating conditions included in the data set were exploited since they are able to excite different types of dynamics. 
Thus, reduced-order models were used to estimate a subset of parameters that can be directly linked to the targeted P2D model, and the P2D model was then used to estimate model parameters that require more excitation and therefore shorter data sets. 
Our analysis showed that some physical parameters for solid and electrolyte phase cannot be uniquely identified from the data. 
Moreover, the identified grouped parameters considering low current measurements and an SPM incurred into twice the error than the one associated to the direct estimation of the P2D model but the results where obtained at least 500 times faster. 
However, if small errors are required, the optimal grouped parameters from the SPM can still be used to initialize the identification of the P2D model, which saves approximately half of the required optimization time. 
Finally, the identification of grouped parameters considering mid to high current measurements and the P2D model is just a fraction (e.g. 6\% on average in the worst case) of the other measurements, which does not make it prohibitive under such conditions.

%\vspace{-0.2cm}

\section*{{APPENDIX}} \label{app}

The standard P2D model is reported in Table \ref{t:p2d_app} for the sake of completeness, together with the description of the associated physical variables in Table \ref{t:phys_params}.

% \addtolength{\textheight}{-12cm}   % This command serves to balance the column lengths
\addtolength{\textheight}{-4cm}

\bibliographystyle{IEEEtran}
\bibliography{IEEEexample}

\end{document}